\documentclass[journal, hidelinks, 10pt]{IEEEtran}
\usepackage{algpseudocode}
\usepackage{amsmath,amsfonts,amssymb}
\usepackage{mathtools}
\usepackage{array}
\usepackage[caption=false, font=normalsize, labelfont=sf, textfont=sf]{subfig}
\usepackage{textcomp}
\usepackage{stfloats}
\usepackage{url}
\usepackage{verbatim}
\usepackage{algorithm}
\usepackage{siunitx}
\usepackage{graphicx}
\graphicspath{{figures/}}
\usepackage{orcidlink}
\usepackage{cite}
\usepackage{xcolor}
\usepackage{fancyhdr}

\usepackage[toc, acronym]{glossaries}
\glsdisablehyper
\def\BibTeX{{\rm B\kern-.05em{\sc i\kern-.025em b}\kern-.08em
    T\kern-.1667em\lower.7ex\hbox{E}\kern-.125emX}}
\usepackage{balance}
\usepackage[final]{microtype}
\usepackage{scalerel}
\usepackage{dsfont}

\newtheorem{theorem}{Theorem}

\DeclareMathOperator*{\argmax}{arg\,max}
\DeclareMathOperator*{\argmin}{arg\,min}
\DeclareMathOperator{\der}{\mathrm{d}\!}

\DeclareMathOperator{\mse}{\mathrm{MSE}}
\DeclareMathOperator{\bias}{\mathrm{b}}

\newcommand{\prob}[2]{\ensuremath{\mathrm{Pr}\ifthenelse{\boolean{#2}}{\left[#1\right]}{[#1]}}}
\newcommand{\expec}[3]{\ensuremath{\mathrm{E}_{#2}\ifthenelse{\boolean{#3}}{\left[#1\right]}{[#1]}}}    
\newcommand{\var}[3]{\ensuremath{\mathrm{var}_{#2}\ifthenelse{\boolean{#3}}{\left(#1\right)}{(#1)}}}    
\newcommand{\trace}[2]{\ensuremath{\mathrm{tr}\ifthenelse{\boolean{#2}}{\left(#1\right)}{(#1)}}}        
\newcommand{\frob}[2]{\ensuremath{\ifthenelse{\boolean{#2}}{\left\|#1\right\|}{\|#1\|}_{\mathrm{F}}}}   
\newcommand{\I}[3]{\ensuremath{\mathrm{I}\ifthenelse{\boolean{#3}}{\left(#1; #2\right)}{(#1; #2)}}}     
\newcommand{\h}[2]{\ensuremath{\mathrm{h}\ifthenelse{\boolean{#2}}{\left(#1\right)}{(#1)}}}             

\newcommand{\cov}[1]{\ensuremath{\mathbf{C}_{#1}}}          
\newcommand{\id}[1]{\ensuremath{\mathbf{I}_{#1}}}           
\newcommand{\bsf}[1]{\ensuremath{\boldsymbol{\mathsf{#1}}}} 
\newcommand{\herm}[1]{\ensuremath{#1^{\mathrm{H}}}}         
\newcommand{\trans}[1]{\ensuremath{#1^{\mathrm{T}}}}        

\newcommand{\real}[1]{\ensuremath{\Re\left(#1\right)}}      
\newcommand{\imag}[1]{\ensuremath{\Im\left(#1\right)}}      

\newcommand{\ie}{\textit{i.e. }}
\newcommand{\eg}{\textit{e.g. }}

\let\originalleft\left
\let\originalright\right
\renewcommand{\left}{\mathopen{}\mathclose\bgroup\originalleft}
\renewcommand{\right}{\aftergroup\egroup\originalright}

\newcommand{\changefont}{
    \color{blue}\fontsize{9}{9}\selectfont
}

\let\oldmaketitle\maketitle
\renewcommand{\maketitle}{%
     \oldmaketitle
     \thispagestyle{fancy}
}

\begin{document}
\newacronym{simo}{SIMO}{single-input multiple-output}
\newacronym{ofdm}{OFDM}{orthogonal frequency division multiplexing}
\newacronym{crb}{CRB}{Cramér--Rao bound}
\newacronym{mvue}{MVUE}{minimum variance unbiased estimator}
\newacronym{bque}{BQUE}{best quadratic unbiased estimator}
\newacronym{mse}{MSE}{mean squared error}
\newacronym{gmm}{GMM}{Gaussian mixture model}
\newacronym{clt}{CLT}{central limit theorem}
\newacronym{bs}{BS}{base station}
\newacronym{mimo}{MIMO}{multiple-input multiple-output}
\newacronym{ml}{ML}{maximum likelihood}
\newacronym{ask}{ASK}{amplitude-shift keying}
\newacronym{blue}{BLUE}{best linear unbiased estimator}
\newacronym{wrt}{w.r.t.}{with respect to}
\newacronym{mmse}{MMSE}{minimum mean squared error}
\newacronym{qmmse}{QMMSE}{quadratic minimum mean squared error}
\newacronym{lmmse}{LMMSE}{linear minimum mean squared error}
\newacronym{qmvue}{QMVUE}{quadratic minimum variance unbiased estimator}
\newacronym{acqe}{ACQE}{averaged covariance quadratic estimator}
\newacronym{ccqe}{CCQD}{corrected covariance quadratic detector}
\newacronym{pam}{PAM}{pulse-amplitude modulation}
\newacronym{snr}{SNR}{signal-to-noise ratio}
\newacronym{pep}{PEP}{pairwise error probability}
\newacronym{llr}{LLR}{log-likelihood ratio}
\newacronym{ook}{OOK}{on-off keying}
\newacronym{ed}{ED}{energy detector}
\newacronym{csi}{CSI}{channel state information}
\newacronym{csir}{CSIR}{channel state information at the receiver}
\newacronym{pdf}{PDF}{probability density function}
\newacronym{ser}{SER}{symbol error rate}
\newacronym{mgf}{MGF}{moment-generating function}
\newacronym{pzi}{PZI}{Paley--Zygmund inequality}
\newacronym{cdf}{CDF}{cumulative distribution function}
\newacronym{ee}{EE}{energy estimator}
\newacronym{mi}{MI}{mutual information}
\newacronym{abque}{ABQUE}{assisted best quadratic unbiased estimator}
\newacronym{iiot}{IIoT}{industrial internet of things}
\newacronym{iot}{IoT}{internet of things}
\newacronym{mmtc}{mMTC}{massive machine-type communication}
\newacronym{urllc}{URLLC}{ultra-reliable and low-latency communication}
\newacronym{im}{IM}{index modulation}
\newacronym{ufc}{UFC}{uniquely factorable constellation}
\newacronym{ber}{BER}{bit error rate}

\title{Quadratic Detection in Noncoherent Massive SIMO Systems over Correlated Channels}
\author{Marc\,Vilà-Insa\,\orcidlink{0000-0002-7032-1411}, \IEEEmembership{Graduate Student Member, IEEE},
	Aniol~Martí\,\orcidlink{0000-0002-5600-8541}, \IEEEmembership{Graduate Student Member, IEEE},\\
	Jaume~Riba\,\orcidlink{0000-0002-5515-8169}, \IEEEmembership{Senior Member, IEEE},
	and Meritxell~Lamarca\,\orcidlink{0000-0002-8067-6435}, \IEEEmembership{Member, IEEE}
	\thanks{This work was (partially) funded by project RODIN (PID2019-105717RB-C22) by MCIU/AEI/10.13039/501100011033, project MAYTE (PID2022-136512OB-C21) by MCIU/AEI/10.13039/501100011033 and ERDF ``A way of making Europe'', grant 2021 SGR 01033 and grant 2022 FI SDUR 00164 by Departament de Recerca i Universitats de la Generalitat de Catalunya and grant 2023 FI ``Joan Oró'' 00050 by Departament de Recerca i Universitats de la Generalitat de Catalunya and the ESF+.}%
	\thanks{The authors are with the Signal Processing and Communications Group (SPCOM), Departament de Teoria del Senyal i Comunicacions, Universitat Politècnica de Catalunya (UPC), 08034 Barcelona, Spain (e-mail: \{marc.vila.insa, aniol.marti, jaume.riba, meritxell.lamarca\}@upc.edu).}
}

\maketitle

\begin{abstract}
    With the goal of enabling ultrareliable and low-latency wireless communications for \acrfull{iiot}, this paper studies the use of energy-based modulations in noncoherent massive \acrfull{simo} systems.
    We consider a one-shot communication over a channel with correlated Rayleigh fading and colored Gaussian noise, in which the receiver has statistical \acrfull{csi}.
    We first provide a theoretical analysis on the limitations of unipolar \acrfull{pam} in systems of this kind, based on \acrlong{ml} detection.
    The existence of a fundamental error floor at high \acrfull{snr} regimes is proved for constellations with more than two energy levels, when no (statistical) \acrshort{csi} is available at the transmitter.
    In the main body of the paper, we present a design framework for quadratic detectors that generalizes the widely-used energy detector, to better exploit the statistical knowledge of the channel.
    This allows us to design receivers optimized according to information-theoretic criteria that exhibit lower error rates at moderate and high \acrshort{snr}.
    We subsequently derive an analytic approximation for the error probability of a general class of quadratic detectors in the large array regime.
    Finally, we numerically validate it and discuss the outage probability of the system.
\end{abstract}


\section{Introduction}
    \IEEEPARstart{I}{ndustrial} internet of things (\acrshort{iiot}) is envisioned as the most promising extension of the already well-established \acrfull{iot}.
    It has been attracting significant attention from both public and private sectors, since it is the key enabler for many potential applications that will impact both business and society as a whole.
    Some examples of such new applications are smart-grid energy management, smart cities, interconnected medical systems and autonomous drones~\cite{Mumtaz2017}.

    In short, \acrshort{iot} is the interconnection and management of intelligent devices with almost no human intervention, mainly focusing on low-power consumer applications.
    On the contrary, its evolved form \acrshort{iiot} is targeted to mission-critical industrial cases~\cite{Mahmood2022}.
    The wireless networks involved in these scenarios are usually characterized by a high density of low-power/low-complexity terminals, the majority of which are inactive most of the time.
    The information being transmitted through these networks mainly consists in control and telemetry data collected by machine sensors.
    Therefore, while the volume of traffic is low, its transmission must be highly reliable and fulfill stringent latency constraints.
    To accomplish these challenging requirements, \acrshort{iiot} leverages two operation modes of the fifth generation (5G) of wireless systems: \acrfull{mmtc} and \acrfull{urllc}~\cite{Vaezi2022}.

    Achieving high reliability over wireless links is challenging due to the instability of the physical medium, caused by fading, interference and other phenomena.
    To combat them, high diversity is of upmost importance, which can be achieved in time, frequency or space~\cite{gao_energy-efficient_2019}.
    Traditionally, time and frequency have been the two main domains from which wireless systems have obtained diversity.
    However, they are not suitable for the development of \acrshort{iiot}.
    On the one hand, time diversity is antagonistic to the low-latency requirements.
    On the other, the scarcity in spectral resources is aggravated by the deployment of massive networks of terminals using frequency diversity.
    Therefore, the most fitting solution is spatial diversity, obtained from the use of arrays with many antennas~\cite{chowdhury_scaling_2016}.
    In particular, \acrfull{simo} architectures seem like a logical choice to implement the communication from machine devices to a central node.
    While most terminals in an \acrshort{iiot} network are power- and complexity-limited, the central node can allocate a large amount of antennas (\ie massive arrays), which results in even more dramatic diversity gains~\cite{Chowdhury2017}.

    The use of large arrays in \acrshort{simo} architectures (\ie massive \acrshort{simo}) entails several benefits over its conventional counterpart.
    For instance, small-scale fading and noise effects decrease as the number of antennas increases.
    This stabilization of channel statistics is known as \textit{channel hardening}~\cite[Ch.~1.3]{marzetta_fundamentals_2016} (comparable to sphere hardening~\cite[Ch.~5.1.2]{Tse2005}).
    To exploit this phenomenon, it is usually required for the receiver to acquire instantaneous \acrfull{csi}, and implement coherent detection of the transmitted data with it.
    The estimation of this \acrshort{csi} is usually performed within a training phase, whose complexity increases with the number of antennas.
    In a regime such as \acrshort{iiot}, the size of training packets can be comparable to the short data payloads transmitted, which is problematic in terms of latency.
    The fraction of resources spent in instantaneous \acrshort{csi} acquisition is aggravated in high-mobility scenarios~\cite{gao_energy-efficient_2019}.
    To combat these limitations, a noncoherent approach can be adopted, in which neither the transmitter nor the receiver have instantaneous \acrshort{csi}.
    Instead, these schemes are able to communicate and exploit channel hardening by using \textit{statistical \acrshort{csi}} (\ie knowledge about the channel and noise distributions), whose acquisition requires simpler training because it changes much more slowly~\cite[Sec.~4.5]{heath_jr_foundations_2018}.
    
    Many authors have studied the problem of noncoherent systems with single-antenna transmitters for \acrshort{urllc} applications~\cite{Bana2018,Han2021,Li2021,Duong2023}, and several of them have envisioned schemes based on energy detection~\cite{Manolakos2016,jing_design_2016,gao_energy-efficient_2019,xie_non-coherent_2020,han_constellation_2022}.
    Most of these works are based on the \acrfull{ed} scheme~\cite[Sec.~5.3]{Kay1998}.
    It arises naturally as a special case of the \acrfull{ml} detector under uncorrelated Rayleigh fading and uses the squared norm of the received signal as its statistic, which is sufficient~\cite{gao_energy-efficient_2019}.
    In this scenario, the \acrshort{ed} is optimal and allows for low-complexity implementations.
    Furthermore, energy-based noncoherent schemes display the same rate scaling behaviour (in terms of receiver antennas) as their coherent counterparts~\cite{chowdhury_scaling_2016}.

    Despite the analytical simplicity entailed by isotropic channels, it is an unrealistic model for fading statistics when using massive arrays~\cite{bjornson_massive_2016}.
    The noncoherent energy-based setting for arbitrary fading remains mostly unexplored in the literature.
    A notable exception is~\cite{han_constellation_2022}, in which the \acrshort{ml} detector is analyzed under correlated fading and an optimized constellation design is developed.
    Beside requiring a much higher computational complexity (compared to \acrshort{ed}), \acrshort{ml} detection in this general setting is mathematically more involved and does not yield tractable error probability expressions.
    Nevertheless, using \acrshort{ed} under nonisotropic fading is sub-optimal and its performance is known to be severely degraded by channel correlation~\cite{Kim2009,Sharma2015,Song2020}.
    The statistic used by \acrshort{ed} is not sufficient under general fading and, therefore, does not fully exploit the knowledge of second-order moments of the channel.

    In this paper, we consider a noncoherent, energy-based massive \acrshort{simo} system in which a single-antenna terminal communicates to a central node equipped with a large array over a correlated Rayleigh fading channel.
    Statistical \acrshort{csi} is available at the receiver but not at the transmitter, which is low-complexity, as is common in many \acrshort{iiot} implementations~\cite{durisi_toward_2016}.
    Information is modulated on the transmitted energy and the receiver decodes it symbol-by-symbol (\ie one-shot scheme), thus a unipolar \acrfull{pam} constellation is adopted.
    Contrary to existing approaches, we propose an architecture for a family of quadratic detectors that can thoroughly exploit statistical \acrshort{csi} (hence generalizing \acrshort{ed}) and whose performance is analytically tractable.
    
    We now provide an outline and a brief breakdown of the body of the work, highlighting its main contributions:
    \begin{itemize}
        \item In Section~\ref{sec:preliminaries}, the problem of interest is formulated and analyzed asymptotically under the well-known \textit{uniquely identifiable constellation} assumption~\cite{Dong2023}.
        For an asymptotically large number of antennas, we prove it is a necessary and sufficient condition for the error probability to vanish (Theorem~\ref{thm:1}).
        This statement is very relevant in the setting of noncoherent reception with massive arrays, as it is a result of channel hardening.
        Moreover, for asymptotically high \acrfull{snr}, we demonstrate that unique identification is insufficient to arbitrarily reduce error probability in constellations with more than two symbols (Theorem~\ref{th:snr_floor}).
        To do so, we prove the existence of a fundamental error floor at high \acrshort{snr}, whose characterization is of upmost importance in the context of energy efficiency for \acrshort{iiot}.
        This result generalizes related ones from~\cite{jing_design_2016}, in which the authors establish a high \acrshort{snr} error floor under isotropic Rayleigh fading and attribute it to channel energy uncertainties.
        Similar phenomena are also observed numerically in~\cite{xie_non-coherent_2020}.
        Note that this issue is not present when the transmitter can exploit statistical \acrshort{csi} by employing optimized constellations that adapt to each \acrshort{snr} level~\cite{gao_energy-efficient_2019}.
        \item In Section~\ref{sec:estimation}, we introduce a general symbol detection framework based on quadratic energy statistics, for which \acrshort{ed} is a simple case.
        Its structure allows to decouple the reception into an estimation phase and a decision one.
        To deal with the first step, we derive the Bayesian \acrfull{qmmse} estimator, which arises from an optimization problem based on information-theoretic criteria.
        As a byproduct, we obtain the \acrfull{bque} for the signal energy at the receiver, and prove it is an unbiased estimator whose variance reaches the \acrfull{crb}.
        We additionally show it is unrealizable, revealing the \acrfull{mvue} does not exist for such scenario.
        \item Section~\ref{sec:detection} deals with the decision step.
        By use of the \acrfull{clt}, we are able to approximate the output distribution of the previous estimator for a large number of antennas.
        This allows us to obtain an algorithm (Algorithm~\ref{alg:gaussian_thresholds}) to find suitable detection thresholds, as well as a simple analytic expression for the error probability, as a sum of $\mathrm{Q}$-functions.
        This section ends with the introduction of a decision-directed detection scheme, the \acrfull{abque}, which leverages the output of the \acrshort{ed} to enable the \acrshort{bque}.
        \item In Section~\ref{sec:numerical_results}, we numerically assess the performance of the presented detectors.
        In particular, we illustrate how channel hardening is better exploited in statistical \acrshort{csi}-aware detectors to reduce outage probability.
        In terms of average \acrfull{ser},
        \acrshort{abque} displays close to \acrshort{ml} performance in most regimes, while the other quadratic detectors greatly outperform \acrshort{ed} under correlated channels.
        Remarkably, the previously derived analytic approximations for error probabilities accurately match these numerical results.
        \item Finally, in Section~\ref{sec:comparison}, we summarize the main properties of the analyzed detectors and compare them qualitatively, in terms of computational complexity, performance at various \acrshort{snr} and correlation regimes, and mathematical tractability.
    \end{itemize}

    The following notation is used throughout the text.
    Boldface lowercase and uppercase letters denote vectors and matrices, respectively.
    Given a matrix $\mathbf{A}$, the element in its row $r$ and column $c$ is denoted as $[\mathbf{A}]_{r,c}$.
    The transpose operator is $\trans{\cdot}$ and the conjugate transpose one is $\herm{\cdot}$.
    The trace operator is $\trace{\cdot}{false}$, the determinant of a matrix $\mathbf{A}$ is given by $|\mathbf{A}|$ and its Frobenius norm is $\frob{\mathbf{A}}{false}\triangleq\sqrt{\trace{\herm{\mathbf{A}}\mathbf{A}}{false}}$.
    The sets of real and complex numbers are $\mathbb{R}$ and $\mathbb{C}$, respectively.
    The imaginary unit is represented with $\mathrm{j}$, and the real and imaginary parts of a complex number are $\real{\cdot}$ and $\imag{\cdot}$.
    Random variables are indicated with sans-serif.
    Expressing $\bsf{a}|b$ is an abuse of notation to represent the conditioning $\bsf{a}|\mathsf{b}=b$.
    Expectation with respect to the distribution of $\mathsf{a}$ is $\expec{\cdot}{\mathsf{a}}{false}$.
    A multivariate circularly-symmetric complex normal vector $\bsf{a}$ with mean $\mathbf{b}$ and covariance matrix $\cov{\bsf{a}}$ is denoted $\bsf{a}\sim\mathcal{CN}(\mathbf{b},\cov{\bsf{a}})$.
    Convergence in distribution is expressed with $\xrightarrow{\mathrm{d}}$.
    Differential entropy is represented with $\h{\cdot}{false}$.
    Finally, a polynomial $p(x)=a_Nx^N+a_{N-1}x^{N-1}+\cdots+a_1x+a_0$ is compactly expressed as $\mathcal{P}(a_0,\dots,a_N)$.

\section{Preliminaries} \label{sec:preliminaries}
    \subsection{Problem formulation}
        Consider a narrowband massive \acrshort{simo} architecture with a single-antenna transmitter and a receiver \acrfull{bs} equipped with $N$ antennas.
        The communication is one-shot and performed through a fast fading channel, modelled as a random variable $\bsf{h}$ that remains constant for a single channel use and changes into an independent realization in the next one.
        The transmitter sends an equiprobable symbol $x\in\mathbb{C}$ selected from a $M$-ary constellation $\mathcal{X}\triangleq\{x_1, \dots, x_M\}$, for $M\geq2$.
        An average transmitted power constraint is assumed (\ie $\expec{|\mathsf{x}|^2}{\mathsf{x}}{false}=1$).
        
        The signal at the receiver is expressed using a complex baseband representation:
        \begin{equation}
            \bsf{y} = \bsf{h}\mathsf{x} + \bsf{z}, \quad \bsf{y}, \bsf{h}, \bsf{z} \in \mathbb{C}^{N},\label{eq:rx_signal}
        \end{equation}
        being $\bsf{z}$ an additive Gaussian noise component.
        The receiver has full statistical \acrshort{csi}\footnote{Acquisition of second-order statistical properties of both channel and noise is a prominent research topic in communications.
        Refer, for example, to~\cite{Sungwoo2018, Wang2019, Lu2024} and references therein for a variety of methods and techniques.}, \ie it is aware of the distributions of both $\bsf{h}$ and $\bsf{z}$, but not their realizations.
        The transmitter is completely unaware of the channel state, \ie there is no \acrshort{csi} at the transmitter (CSIT).
        The fading is assumed correlated Rayleigh: $\bsf{h}\sim\mathcal{CN}(\mathbf{0}_{N},\cov{\bsf{h}})$.
        This is consistent with general, well-established channel models~\cite[Ch.~3]{heath_jr_foundations_2018}, as well as state-of-the-art ones, such as those involved in near-field communications~\cite{liu_near-field_2023}.
        Similarly, the noise is distributed as $\bsf{z}\sim\mathcal{CN}(\mathbf{0}_{N},\cov{\bsf{z}})$ with arbitrary correlation, which accounts for both colored noise and multiuser interference.
        The average \acrshort{snr} at the receiver is defined as follows:
        \begin{equation}
            \alpha\triangleq\frac{\expec{|\mathsf{x}|^2\|\bsf{h}\|^2}{\mathsf{x},\bsf{h}}{false}}{\expec{\|\bsf{z}\|^2}{\bsf{z}}{false}}=\frac{\expec{|\mathsf{x}|^2}{\mathsf{x}}{false}\trace{\expec{\bsf{h}\herm{\bsf{h}}}{\bsf{h}}{false}}{false}}{\trace{\expec{\bsf{z}\herm{\bsf{z}}}{\bsf{z}}{false}}{false}}=\frac{\trace{\cov{\bsf{h}}}{false}}{\trace{\cov{\bsf{z}}}{false}},
        \end{equation}
        which has been derived using independence between $\bsf{h}$ and $\mathsf{x}$, the cyclic property of the trace and linearity of the expectation.

    \subsection{ML detector}
        For a constellation $\mathcal{X}$ with equiprobable symbols, the probability of error is defined as
        \begin{equation}
            \mathrm{P}_{\epsilon}\triangleq\frac{1}{M}\sum_{x\in\mathcal{X}}\prob{\widehat{x}(\bsf{y})\neq x|\mathsf{x}=x}{true},\label{eq:err_prob}
        \end{equation}
        where $\widehat{x}(\bsf{y})$ is the output of a symbol detector applied to the received signal $\bsf{y}$.
        The receiver that minimizes $\mathrm{P}_{\epsilon}$ is the \acrshort{ml} detector~\cite[Sec.~4.1-1]{proakis_digital_2008}.
        
        The likelihood function of the received signal~\eqref{eq:rx_signal} for a transmitted symbol and given a channel realization is
        \begin{equation}
            f_{\bsf{y}|x,\mathbf{h}}\left(\mathbf{y}\right)=\frac{\exp(-\herm{(\mathbf{y}-\mathbf{h}x)}\cov{\bsf{z}}^{-1}(\mathbf{y}-\mathbf{h}x))}{\pi^{N}|\cov{\bsf{z}}|}.
        \end{equation}
        The channel realization is unknown at the receiver and is removed by marginalizing the previous function (\ie \textit{unconditional model}~\cite{stoica_performance_1990}).
        This results in the following likelihood function:
        \begin{equation}
            f_{\bsf{y}|x}\left(\mathbf{y}\right)=\expec{f_{\bsf{y}|x,\mathbf{h}}\left(\mathbf{y}\right)}{\bsf{h}}{true}=\frac{\exp(-\herm{\mathbf{y}}\cov{\bsf{y}|x}^{-1}\mathbf{y})}{\pi^{N}|\cov{\bsf{y}|x}|},\label{eq:likelihood}
        \end{equation}
        where $\cov{\bsf{y}|x}\triangleq|x|^2\cov{\bsf{h}}+\cov{\bsf{z}}$ is the covariance matrix of the received signal for a given $x$.

        The \acrshort{ml} detector is obtained by maximizing the likelihood function over all possible symbols in $\mathcal{X}$:
        \begin{equation}
            \begin{aligned}
                \widehat{x}_{\mathrm{ML}}&=\argmax_{x\in\mathcal{X}}f_{\bsf{y}|x}\left(\mathbf{y}\right)=\argmax_{x\in\mathcal{X}}\log f_{\bsf{y}|x}\left(\mathbf{y}\right)\\
                &=\argmin_{x\in\mathcal{X}}\herm{\mathbf{y}}\cov{\bsf{y}|x}^{-1}\mathbf{y}+\log|\cov{\bsf{y}|x}|.
            \end{aligned}\label{eq:ml_det}
        \end{equation}

        It is clear from~\eqref{eq:likelihood} and~\eqref{eq:ml_det} that the phase information of $x$ cannot be retrieved from $\mathbf{y}$, since the noncoherent \acrshort{ml} detector only perceives its energy $\varepsilon_i\triangleq|x_i|^2$. For this reason, $\mathcal{X}$ will be constructed from a unipolar \acrfull{pam}:
        \begin{equation}
            \mathcal{X}=\{\sqrt{\varepsilon_1}\triangleq0<\sqrt{\varepsilon_2}<\dots<\sqrt{\varepsilon_M}\}.\label{eq:constellation}
        \end{equation}
        In various related scenarios, it has been proven that the capacity-achieving input distribution is composed of a finite number of discrete mass points, one of which is found at the origin~\cite{AbouFaycal2001,Gursoy2005a}.
        Hence, we set the first symbol in our constellation at 0, in the same manner as in most other works on the topic~\cite{Manolakos2016,chowdhury_scaling_2016,jing_design_2016,han_constellation_2022}.

        We might express
        \begin{equation}
            \cov{\bsf{y}|x}=\cov{\bsf{z}}^{\frac{1}{2}}(|x|^2\cov{\bsf{z}}^{-\frac{1}{2}}\cov{\bsf{h}}\cov{\bsf{z}}^{-\frac{1}{2}}+\id{N})\cov{\bsf{z}}^{\frac{1}{2}},
        \end{equation}
        and then eigendecompose $\cov{\bsf{z}}^{-\frac{1}{2}}\cov{\bsf{h}}\cov{\bsf{z}}^{-\frac{1}{2}}$ as $\mathbf{U}\boldsymbol{\Gamma}\herm{\mathbf{U}}$.
        The diagonal matrix $\boldsymbol{\Gamma}$ contains the spectrum $\{\gamma_n\}_{1\leq n\leq N}$ and $\mathbf{U}$ is its eigenbasis.
        Without loss of generality, both $\cov{\bsf{h}}$ and $\cov{\bsf{z}}$ are assumed full-rank.
        With these transformations, we can define
        \begin{equation}
            \mathbf{r}\triangleq\herm{\mathbf{U}}\cov{\bsf{z}}^{-\frac{1}{2}}\mathbf{y}\Longrightarrow\bsf{r}|x\sim\mathcal{CN}(\mathbf{0}_{N},\cov{\bsf{r}|x}),\label{eq:uncorr}
        \end{equation}
        with $\cov{\bsf{r}|x}\triangleq|x|^2\boldsymbol{\Gamma}+\id{N}$.
        It is obtained by noise whitening and subsequent decorrelation of the received signal $\mathbf{y}$.
        Notice how there is a one-to-one correspondence between $\mathbf{y}$ and $\mathbf{r}$, thus the \acrshort{ml} detector~\eqref{eq:ml_det} can be equivalently expressed as
        \begin{equation}
            \begin{aligned}
                \widehat{x}_{\mathrm{ML}}&=\argmin_{x\in\mathcal{X}}\herm{\mathbf{r}}\cov{\bsf{r}|x}^{-1}\mathbf{r}+\log\left|\cov{\bsf{r}|x}\right|\\
                &=\argmin_{x\in\mathcal{X}}\sum_{n=1}^{N}\frac{|r_n|^2}{|x|^2\gamma_n+1}+\log\left(|x|^2\gamma_n+1\right),
            \end{aligned}\label{eq:ml_decorr}
        \end{equation}
        with $r_n\triangleq[\mathbf{r}]_n$.
        This detector coincides with the one presented in~\cite{han_constellation_2022}.
        From an algebraic perspective, it is more convenient to treat~\eqref{eq:ml_decorr} rather than~\eqref{eq:ml_det} in subsequent derivations, since $\cov{\bsf{r}|x}$ is diagonal.
        Therefore, we will refer to $\bsf{r}|x$ instead of $\bsf{y}|x$ without loss of generality, even though the analysis is valid for a receiver operating directly on $\bsf{y}|x$.

    \subsection{Asymptotic regimes}
        It is of theoretical interest to analyze the performance of the presented communication system in various asymptotic regimes, namely as the number of receiving antennas or \acrshort{snr} grow without bound.
        Let
        \begin{equation}
            \begin{aligned}
                \mathrm{P}_{a\to b}&\triangleq\prob{\widehat{x}=x_b|\mathsf{x}=x_a}{false}\\
                &=\prob{f_{\bsf{r}|x_a}(\bsf{r})\leq f_{\bsf{r}|x_b}(\bsf{r})\big|\mathsf{x}=x_a}{true}\\
                &=\prob{\mathrm{L}_{a,b}(\bsf{r})\leq0\big|\mathsf{x}=x_a}{true}
            \end{aligned}\label{eq:pep}
        \end{equation}
        be the \acrfull{pep} associated with transmitting $x_a$ and detecting $x_b$ at the receiver, in which
        \begin{equation}
            \begin{aligned}
                \mathrm{L}_{a,b}\left(\mathbf{r}\right)&\triangleq\log\frac{f_{\bsf{r}|x_a}\left(\mathbf{r}\right)}{f_{\bsf{r}|x_b}\left(\mathbf{r}\right)}\\
                &=\herm{\mathbf{r}}\left(\cov{\bsf{r}|x_b}^{-1}-\cov{\bsf{r}|x_a}^{-1}\right)\mathbf{r}+\log\frac{|\cov{\bsf{r}|x_b}|}{|\cov{\bsf{r}|x_a}|}
            \end{aligned}\label{eq:llr}
        \end{equation}
        is the \acrfull{llr}~\cite[Ch.~3]{levy_principles_2008} between hypotheses $a$ and $b$.
        With the maximum \acrshort{pep} of the constellation, we can bound the error probability presented in~\eqref{eq:err_prob}, as stated in~\cite[Sec.~4.2-3]{proakis_digital_2008}:
        \begin{equation}
            \max_{x_a\neq x_b\in\mathcal{X}}\frac{1}{M}\mathrm{P}_{a\to b}\leq\mathrm{P}_{\epsilon}\leq\max_{x_a\neq x_b\in\mathcal{X}}\left(M-1\right)\mathrm{P}_{a\to b}.
        \end{equation}
        This implies that the error probability will vanish if and only if the maximum \acrshort{pep} does as well:
        \begin{equation}
            \begin{aligned}
                \lim_{N\to\infty}\max_{x_a\neq x_b\in\mathcal{X}}\mathrm{P}_{a\to b}=0&\iff\lim_{N\to\infty}\mathrm{P}_{\epsilon}=0\\
                \lim_{\alpha\to\infty}\max_{x_a\neq x_b\in\mathcal{X}}\mathrm{P}_{a\to b}=0&\iff\lim_{\alpha\to\infty}\mathrm{P}_{\epsilon}=0.
            \end{aligned}
        \end{equation}

        We now state two fundamental results regarding the performance of the presented system under asymptotic regimes.
        \begin{theorem}\label{thm:1}
            Let $\mathrm{\Theta}(N)$ be the number of nonzero eigenvalues of $\mathbf{\Gamma}$ such that it grows without bound for $N\to\infty$.
            The following condition is \textit{necessary and sufficient} for the error probability of a constellation $\mathcal{X}$ to vanish for an increasing number of receiving antennas:
            \begin{equation}
                |x_a|^2\neq|x_b|^2 \iff x_a\neq x_b,\quad\forall x_a,x_b\in\mathcal{X}.\label{eq:puf}
            \end{equation}
        \end{theorem}
        \begin{IEEEproof}
            See Appendix \ref{ap:thm1}.
        \end{IEEEproof}

\begin{figure}[ht]
            \centering
            \includegraphics[width=\columnwidth]{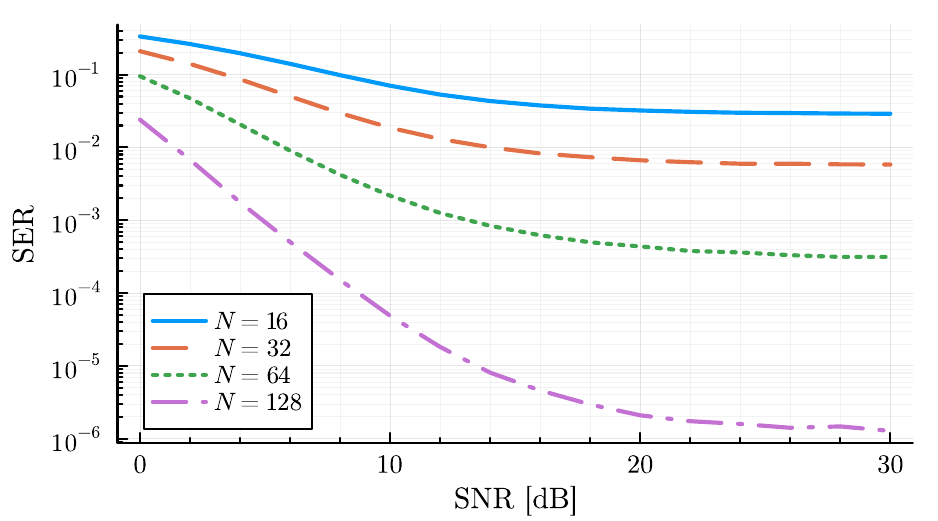}
            \caption{Monte Carlo results of the \acrshort{ser} for the \acrshort{ml} detector in terms of \acrshort{snr}.
            Various numbers of antennas have been considered under a correlated Rayleigh channel with $\rho=0.8$ with a uniform unipolar 4-\acrshort{ask} modulation (see Section~\ref{sec:numerical_results} for a detailed channel model description).}
            \label{fig:ml_error}
        \end{figure}

        \begin{theorem}
            \label{th:snr_floor}
            For a finite number of receiving antennas, the error probability of a constellation of type~\eqref{eq:constellation} vanishes for increasing \acrshort{snr} if and only if $M=2$.
        \end{theorem}
        \begin{IEEEproof}
            See Appendix \ref{ap:thm2}.
        \end{IEEEproof}
        What these results imply is that communication with the system considered in this paper is asymptotically error-free for an increasing number of receiving antennas but not for increasing \acrshort{snr}.
        This behavior is illustrated in Fig.~\ref{fig:ml_error}.
        
    \subsection{Energy detection and high SNR approximation}\label{ssec:ed}
        There is a special case of the \acrshort{ml} problem presented in~\eqref{eq:ml_det} which has been widely studied in the literature~\cite{chowdhury_scaling_2016,gao_energy-efficient_2019}: the \textit{isotropic channel}.
        Under this model, the spectral matrix $\boldsymbol{\Gamma}$ is proportional to the identity (\ie $\boldsymbol{\Gamma}\triangleq\alpha\id{N}$).
        This assumption greatly simplifies \eqref{eq:ml_decorr}, since $\bsf{r}|x\sim\mathcal{CN}(\mathbf{0}_{N},(|x|^2\alpha+1)\id{N})$:
        \begin{equation}
            \widehat{x}_{\mathrm{ML}}^{(\mathrm{iso})}=\arg\min_{x\in\mathcal{X}}\frac{\left\|\mathbf{r}\right\|^2}{|x|^2\alpha+1}+N\log\left(|x|^2\alpha+1\right).\label{eq:ml_ed}
        \end{equation}
        In this scenario, $\|\mathbf{r}\|^2$ is a sufficient statistic~\cite{chowdhury_scaling_2016}.

        Another relevant simplification of the \acrshort{ml} detector that results in a similar expression emerges at high \acrshort{snr}.
        Given a constellation $\mathcal{X}'$ that does not include the null symbol $x=0$, the covariance matrix of the received signal reduces to
        \begin{equation}
            \lim_{\alpha\to\infty}\cov{\bsf{r}|x}=\lim_{\alpha\to\infty}|x|^2\mathbf{\Gamma},
        \end{equation}
        with which detector~\eqref{eq:ml_decorr} simplifies to
        \begin{equation}
            \lim_{\alpha\to\infty}\widehat{x}_{\mathrm{ML}}=\lim_{\alpha\to\infty}\argmin_{x\in\mathcal{X}'}\frac{\herm{\mathbf{r}}\mathbf{\Gamma}^{-1}\mathbf{r}}{|x|^2}+N\log\left(|x|^2\right)\label{eq:ml_hsnr}
        \end{equation}
        and $\herm{\mathbf{r}}\mathbf{\Gamma}^{-1}\mathbf{r}$ becomes a sufficient statistic.
        This result is no longer valid for the constellation considered in~\eqref{eq:constellation}, as $\cov{\bsf{r}|x=0}=\id{N}$, so this symbol has to be treated separately to perform \acrshort{ml} detection.

        In both special cases~\eqref{eq:ml_ed} and~\eqref{eq:ml_hsnr}, the \acrshort{ml} decoder can be decoupled into a two-stage process:
        \begin{enumerate}
            \item The computation of a quadratic statistic.
            \item A one-dimensional decision problem that detects the transmitted symbol solely using the proposed statistic.
        \end{enumerate}
        Motivated by this observation, a generalization of this two-step approach for arbitrary channel and noise spectra is proposed in the sequel.
        Sections~\ref{sec:estimation} and~\ref{sec:detection} address the design of the first and second steps of this procedure.   

\section{Energy statistic} \label{sec:estimation}
    The first simplified receiver described in Section~\ref{ssec:ed}, known as \acrshort{ed}, is widely used in the literature due to its low complexity and optimality within the isotropic channel.
    Nevertheless, when the channel is arbitrarily correlated, the \acrshort{ed} does not fully exploit the statistical \acrshort{csi} and is no longer optimal.
    Therefore, in this paper, we present a family of one-dimensional detectors that incorporate channel correlation.
    We have chosen a quadratic structure (in data) that generalizes the one for \acrshort{ed} and is adequate when dealing with variables related to second-order moments (such as the transmitted energy $|x|^2$, in which we convey information):
    \begin{equation}
        \widehat{\varepsilon}\left(\mathbf{r}\right) \triangleq \herm{\mathbf{r}} \mathbf{A} \mathbf{r} + c.\label{eq:full_quad}
    \end{equation}
    Note that no linear term is considered because $\expec{\bsf{r}|x}{}{false}=\mathbf{0}_N$.
    Its design is split into two phases:
    \begin{enumerate}
        \item Perform the energy estimation with~\eqref{eq:full_quad}.
        \item Detect the transmitted symbol by classifying the estimate according to some detection regions.
    \end{enumerate}

    Throughout this section, we consider the received signal from~\eqref{eq:uncorr}, $\bsf{r}|\varepsilon\sim\mathcal{CN}(\mathbf{0}_N,\cov{\bsf{r}|\varepsilon}\triangleq\varepsilon\mathbf{\Gamma}+\id{N})$, with transmitted energy $\varepsilon=|x|^2$.

    \subsection{Information-theoretic design criteria} \label{sec:information-criteria}
        The detector based on expression \eqref{eq:full_quad} is suboptimal for most scenarios because $\widehat{\varepsilon}$ is not a sufficient statistic in the general case.
        Nevertheless, we wish to design the coefficients of the quadratic estimator so that it preserves as much information as possible on the transmitted symbol, thus minimizing the error rate performance penalty.
        Hence, the mutual information arises as a natural choice for the design criterion for the coefficients of the quadratic estimator.

        We propose to design coefficients $\mathbf{A}$ and $c$ to maximize the \acrfull{mi} between the transmitted symbol and the estimator output, \ie
        \begin{equation}
            \I{\varepsilon}{\widehat{\varepsilon}}{true} = \h{\varepsilon}{true} - \h{\varepsilon|\widehat{\varepsilon}}{true}\implies\widehat{\varepsilon}_{\mathrm{OPT}}\triangleq\arg\max_{\widehat{\varepsilon}\in\mathcal{Q}}\I{\varepsilon}{\widehat{\varepsilon}}{true},
        \end{equation}
        where $\mathcal{Q}$ is the set of all estimators of the form~\eqref{eq:full_quad}.
        This problem is, in general, very hard to solve analytically.
        Motivated by~\cite{pastore_capacity_2018}, we obtain a lower bound on the \acrshort{mi}:
        \begin{align}
            \I{\varepsilon}{\widehat{\varepsilon}}{true} &= \h{\varepsilon}{true} - \h{\varepsilon-\widehat{\varepsilon}|\widehat{\varepsilon}}{true}\label{eq:translation}\\
            &\geq\h{\varepsilon}{true} - \h{\varepsilon-\widehat{\varepsilon}}{true}\label{eq:condition}.
        \end{align}
        Equality~\eqref{eq:translation} is due to the translation invariance of differential entropy~\cite[Th.~8.6.3]{cover_elements_2005}, while inequality~\eqref{eq:condition} is due to~\cite[Th.~8.6.1]{cover_elements_2005}.
        If $\xi\triangleq\varepsilon-\widehat{\varepsilon}$ is interpreted as the estimation error, this lower bound will become tighter the less $\xi$ depends on $\widehat{\varepsilon}$, \ie the better the estimation becomes.

        By using the maximum value of $\h{\xi}{false}$, which is achieved when $\xi$ is Gaussian~\cite[Th.~8.6.5]{cover_elements_2005}, an even simpler lower bound is derived\footnote{In Section~\ref{sec:detection_regions}, we prove the test statistic $\widehat{\varepsilon}|x$ is increasingly Gaussian for large $N$, thus the bound is expected to be asymptotically tight.}:
        \begin{equation} \label{eq:mi_bound}
            \I{\varepsilon}{\widehat{\varepsilon}}{true}\geq\h{\varepsilon}{true}-\frac{1}{2}\left(1+\log\left(2\pi\var{\xi}{}{false}\right)\right)\triangleq\mathrm{I}_{\mathrm{LOW}}.
        \end{equation}
        Therefore, the proposed design criterion simplifies to
        \begin{equation}
            \widehat{\varepsilon}=\arg\max_{\widehat{\varepsilon}\in\mathcal{Q}}\mathrm{I}_{\mathrm{LOW}}\equiv\arg\min_{\widehat{\varepsilon}\in\mathcal{Q}}\var{\xi}{}{false}.\label{eq:info_low}
        \end{equation}
        This variance can be obtained from the average \acrshort{mse} and bias of $\widehat{\varepsilon}$, by applying the \textit{law of total expectation}~\cite[Th.~9.1.5]{blitzstein_introduction_2015}:
        \begin{equation}
            \begin{aligned}
                \var{\xi}{}{false}&=\mse\left(\widehat{\varepsilon}\right)-\bias^2\left(\widehat{\varepsilon}\right)\\
                &=\expec{\expec{\left(\varepsilon-\widehat{\varepsilon}\right)^2}{\bsf{r}|\varepsilon}{true}}{\varepsilon}{true}-\expec{\varepsilon-\expec{\widehat{\varepsilon}}{\bsf{r}|\varepsilon}{true}}{\varepsilon}{true}^2.
            \end{aligned}\label{eq:var_im}
        \end{equation}

    \subsection{Best quadratic unbiased estimator (BQUE)}\label{ssec:bque}
        Before dealing with the full optimization problem, it is of conceptual interest to study a simpler version.
        Consider a \textit{genie-aided} estimator that aims at minimizing $\var{\xi}{}{false}$ but is aware of the exact transmitted energy $\varepsilon$.
        In such a scenario,~\eqref{eq:var_im} reduces to:
        \begin{equation}
            \begin{aligned}
                \var{\xi}{}{false}&=\expec{(\varepsilon-\widehat{\varepsilon})^2}{\bsf{r}|\varepsilon}{true}-\left(\varepsilon-\expec{\widehat{\varepsilon}}{\bsf{r}|\varepsilon}{true}\right)^2\\
                &=\expec{\widehat{\varepsilon}^2}{\bsf{r}|\varepsilon}{true}-\expec{\widehat{\varepsilon}}{\bsf{r}|\varepsilon}{true}^2\equiv\var{\widehat{\varepsilon}|\varepsilon}{}{false}.
            \end{aligned}\label{eq:cond_var}
        \end{equation}
        It depends on the conditional first and second-order moments of $\widehat{\varepsilon}$, which have been derived in Appendix~\ref{ap:quadratic_form_variance}:
        \begin{equation}
            \begin{aligned}
                \expec{\widehat{\varepsilon}}{\bsf{r}|\varepsilon}{true}&=\trace{\mathbf{A}\cov{\bsf{r}|\varepsilon}}{false}+c\\
                \expec{\widehat{\varepsilon}^2}{\bsf{r}|\varepsilon}{true}&=\trace{\mathbf{A}^2\cov{\bsf{r}|\varepsilon}^2}{false}+\left(\trace{\mathbf{A}\cov{\bsf{r}|\varepsilon}}{false} + c\right)^2.
            \end{aligned}\label{eq:moments}
        \end{equation}
        Using them in~\eqref{eq:cond_var}, the conditional variance of $\widehat{\varepsilon}$ becomes
        \begin{equation}
            \var{\widehat{\varepsilon}|\varepsilon}{}{false}=\frob{\mathbf{A}\cov{\bsf{r}|\varepsilon}}{false}^2.
        \end{equation}

        As the affine term $c$ does not affect the variance we want to optimize, we can freely set it.
        For convenience, we propose to choose it to obtain a \textit{conditionally unbiased} estimator, which translates into
        \begin{equation}
            \begin{aligned}
                \expec{\widehat{\varepsilon}}{\bsf{r}|\varepsilon}{true}&=\trace{\mathbf{A}(\varepsilon\mathbf{\Gamma}+\id{N})}{false}+c=\varepsilon\\
                &\implies\begin{cases}
                    c = -\trace{\mathbf{A}}{true}\\
                    \trace{\mathbf{A}\mathbf{\Gamma}}{true} = 1.
                \end{cases}
            \end{aligned}\label{eq:bque_matrix_constraints}
        \end{equation}
        Under conditional unbiasedness, $\var{\widehat{\varepsilon}|\varepsilon}{}{false}$ is equivalent to the conditional \acrshort{mse}, \ie $\mse(\widehat{\varepsilon}|\varepsilon)$.
        Following a similar reasoning to the one behind the \acrfull{blue}~\cite[Ch.~6]{kay_fundamentals_1993}, with the quadratic structure presented, we will term the estimator that minimizes $\mse(\widehat{\varepsilon}|\varepsilon)$ \textit{\acrfull{bque}}\footnote{In~\cite{villares_best_2001}, the authors propose the same name for an estimator with similar structural constraints.}.
        
        Accounting for the constraints in~\eqref{eq:bque_matrix_constraints}, we can solve the optimization problem using the method of Lagrange multipliers:
        \begin{equation}
            \mathcal{L}(\mathbf{A}, \lambda) \triangleq \frob{\mathbf{A}\cov{\bsf{r}|\varepsilon}}{false}^2 + \lambda\left(\trace{\mathbf{A}\mathbf{\Gamma}}{true}-1\right).
        \end{equation}
        We differentiate $\mathcal{L}$ with respect to $\mathbf{A}$ and equate it to $\mathbf{0}$:
        \begin{equation} 
            \begin{aligned}
                \frac{\partial \mathcal{L}(\mathbf{A}, \lambda)}{\partial\mathbf{A}} &= 2\cov{\bsf{r}|\varepsilon}\mathbf{A}\cov{\bsf{r}|\varepsilon} + \lambda\mathbf{\Gamma} = \mathbf{0}_{N\times N}\\
                &\implies \mathbf{A} = -\frac{\lambda}{2}\cov{\bsf{r}|\varepsilon}^{-1}\mathbf{\Gamma}\cov{\bsf{r}|\varepsilon}^{-1}.
            \end{aligned}\label{eq:bque_matrix_lagrange}
        \end{equation}
        Taking into account that $\mathbf{\Gamma}$ and $\cov{\bsf{r}|\varepsilon}$ are diagonal matrices, $\mathbf{A}$ must also be diagonal.
        Enforcing the unbiasedness constraint leaves us with:
        \begin{equation}
            \begin{aligned}
                \trace{\mathbf{A}\mathbf{\Gamma}}{true} &= -\frac{\lambda}{2}\frob{\mathbf{\Gamma}\cov{\bsf{r}|\varepsilon}^{-1}}{false}^2 = 1\\
                &\implies -\frac{\lambda}{2} = \frac{1}{\frob{\mathbf{\Gamma}\cov{\bsf{r}|\varepsilon}^{-1}}{false}^2}.
            \end{aligned}
        \end{equation}
        Substituting in~\eqref{eq:bque_matrix_lagrange} we obtain the matrix of the quadratic form:
        \begin{equation} \label{eq:bque_matrix}
            \mathbf{A}_{\mathrm{BQUE}} = \frac{\mathbf{\Gamma}\cov{\bsf{r}|\varepsilon}^{-2}}{\frob{\mathbf{\Gamma}\cov{\bsf{r}|\varepsilon}^{-1}}{false}^2}.
        \end{equation}
        Finally, the \acrshort{bque} is given by
        \begin{equation} \label{eq:bque}
            \widehat{\varepsilon}_{\mathrm{BQUE}}(\mathbf{r}) = \frac{\herm{\mathbf{r}}\mathbf{\Gamma}\cov{\bsf{r}|\varepsilon}^{-2}\mathbf{r} - \trace{\mathbf{\Gamma}\cov{\bsf{r}|\varepsilon}^{-2}}{false}}{\frob{\mathbf{\Gamma}\cov{\bsf{r}|\varepsilon}^{-1}}{false}^2}.
        \end{equation}
        If we analyze the conditional \acrshort{mse} of this estimator, \ie
        \begin{equation}
            \begin{aligned}
                \mse(\widehat{\varepsilon}_{\mathrm{BQUE}}|\varepsilon)&=\frob{\mathbf{A}_{\mathrm{BQUE}}\cov{\bsf{r}|\varepsilon}}{false}^2\\
                &= \frac{\frob{\mathbf{\Gamma}\cov{\bsf{r}|\varepsilon}^{-1}}{false}^2}{\frob{\mathbf{\Gamma}\cov{\bsf{r}|\varepsilon}^{-1}}{false}^4}=\frac{1}{\frob{\mathbf{\Gamma}\cov{\bsf{r}|\varepsilon}^{-1}}{false}^2},
            \end{aligned}
        \end{equation}
        we observe it coincides with the \acrshort{crb} associated with the estimation of $\varepsilon$, which has been derived in Appendix~\ref{ap:crb}.
        Although this condition is sufficient to state if an estimator is efficient, the \acrshort{bque} depends on the parameter to be estimated, making it unrealizable.
        Therefore, it is not the \acrshort{mvue}~\cite[Ch.~2]{kay_fundamentals_1993}.

    \subsection{QMMSE estimator}
        Developing the \acrshort{bque} has provided us with valuable insights on problem~\eqref{eq:info_low} from the perspective of classical estimation theory.
        We now derive the structure of the quadratic estimator that maximizes $\mathrm{I}_{\mathrm{LOW}}$.
        Recall~\eqref{eq:var_im}, which depends on the average \acrshort{mse} and squared mean bias of $\widehat{\varepsilon}$:
        \begin{equation}
            \begin{aligned}
                \mse\left(\widehat{\varepsilon}\right)&=\expec{\varepsilon^2-2\varepsilon\expec{\widehat{\varepsilon}}{\bsf{r}|\varepsilon}{true}+\expec{\widehat{\varepsilon}^2}{\bsf{r}|\varepsilon}{true}}{\varepsilon}{true}\\
                \bias^2\left(\widehat{\varepsilon}\right)&=1-2\expec{\varepsilon\expec{\widehat{\varepsilon}}{\bsf{r}|\varepsilon}{true}}{\varepsilon}{true}+\expec{\expec{\widehat{\varepsilon}}{\bsf{r}|\varepsilon}{true}}{\varepsilon}{true}^2.
            \end{aligned}
        \end{equation}
        Plugging the first and second-order moments from~\eqref{eq:moments} in the previous expressions yields
        \begin{equation}
            \begin{aligned}
                \mse\left(\widehat{\varepsilon}\right)&=\left(c+\trace{\mathbf{A}\overline{\mathbf{C}}}{false}-1\right)^2+\sigma_{\varepsilon}^2\left(1-\trace{\mathbf{A}\mathbf{\Gamma}}{false}\right)^2\\
                &\quad+\trace{\mathbf{A}^2\mathring{\mathbf{C}}^2}{false}\\
                \bias^2\left(\widehat{\varepsilon}\right)&=\left(c+\trace{\mathbf{A}\overline{\mathbf{C}}}{false}-1\right)^2,
            \end{aligned}
        \end{equation}
        with $\overline{\mathbf{C}}\triangleq\mathbf{\Gamma}+\id{N}$, $\mathring{\mathbf{C}}^2\triangleq(\sigma_{\varepsilon}^2+1)\mathbf{\Gamma}^2+2\mathbf{\Gamma}+\id{N}$ and $\sigma_{\varepsilon}^2\triangleq\var{\varepsilon}{}{false}$.
        Therefore, the variance to be minimized results in
        \begin{equation}
            \var{\xi}{}{false}=\sigma_{\varepsilon}^2\left(1-\trace{\mathbf{A}\mathbf{\Gamma}}{false}\right)^2+\frob{\mathbf{A}\mathring{\mathbf{C}}}{false}^2,\label{eq:var_qmmse}
        \end{equation}
        which is not affected by the value of $c$, similar to what occurred with the \acrshort{bque}.
        If we set it to $c\triangleq1-\trace{\mathbf{A}\overline{\mathbf{C}}}{false}$, the squared mean bias term cancels (\ie $\bias^2(\widehat{\varepsilon})=0$).
        In this situation, the criterion of maximum $\mathrm{I}_{\mathrm{LOW}}$ is equivalent to that of minimum \acrshort{mse} on average.
        This results in the \textit{(Bayesian) \acrshort{qmmse}} estimator, which is an extension of the well-known \acrfull{lmmse} estimator~\cite[Ch.~12]{kay_fundamentals_1993}.

        To obtain its expression, we shall proceed as in Section~\ref{ssec:bque}, by differentiating~\eqref{eq:var_qmmse} and equating it to $\mathbf{0}$:
        \begin{equation}
            \frac{\partial\var{\xi}{}{false}}{\partial\mathbf{A}}=-2\sigma_{\varepsilon}^2\left(1-\trace{\mathbf{A}\mathbf{\Gamma}}{false}\right)\mathbf{\Gamma}+2\mathbf{A}\mathring{\mathbf{C}}^2=\mathbf{0}_{N\times N}.
        \end{equation}
        By isolating $\mathbf{A}$, multiplying it by $\mathbf{\Gamma}$ and computing the trace, we obtain the following term:
        \begin{equation}
            \trace{\mathbf{A\Gamma}}{false}=\frac{\sigma_{\varepsilon}^2\frob{\mathbf{\Gamma}\mathring{\mathbf{C}}^{-1}}{false}^2}{1+\sigma_{\varepsilon}^2\frob{\mathbf{\Gamma}\mathring{\mathbf{C}}^{-1}}{false}^2},
        \end{equation}
        which yields the matrix
        \begin{equation}
            \mathbf{A}_{\mathrm{QMMSE}}=\frac{\sigma_{\varepsilon}^2\mathbf{\Gamma}\mathring{\mathbf{C}}^{-2}}{1+\sigma_{\varepsilon}^2\frob{\mathbf{\Gamma}\mathring{\mathbf{C}}^{-1}}{false}^2}
        \end{equation}
        and the estimator
        \begin{equation}
            \widehat{\varepsilon}_{\mathrm{QMMSE}}(\mathbf{r})=\frac{\sigma_{\varepsilon}^2\herm{\mathbf{r}}\mathbf{\Gamma}\mathring{\mathbf{C}}^{-2}\mathbf{r}+1-\sigma_{\varepsilon}^2\trace{\mathbf{\Gamma}\mathring{\mathbf{C}}^{-2}}{false}}{1+\sigma_{\varepsilon}^2\frob{\mathbf{\Gamma}\mathring{\mathbf{C}}^{-1}}{false}^2}.
        \end{equation}
        The mean \acrshort{mse} value it reaches is
        \begin{equation}
            \mse(\widehat{\varepsilon}_{\mathrm{QMMSE}})=\frac{\sigma_{\varepsilon}^2}{1+\sigma_{\varepsilon}^2\frob{\mathbf{\Gamma}\mathring{\mathbf{C}}^{-1}}{false}^2}.
        \end{equation}

    \subsection{Unified framework for quadratic detectors}
        As stated in Section~\ref{ssec:ed}, the framework outlined in~\eqref{eq:full_quad} is general enough to encapsulate a variety of energy estimators as a first step in symbol detection.
        For instance, an estimator of the form
        \begin{equation}
            \label{eq:energy_estimator}
            \mathbf{A}_{\mathrm{ED}}\triangleq\frac{\id{N}}{\trace{\mathbf{\Gamma}}{true}}\implies\widehat{\varepsilon}_{\mathrm{ED}}(\mathbf{r})=\frac{\|\mathbf{r}\|^2-N}{\trace{\mathbf{\Gamma}}{true}}
        \end{equation}
        with a posterior classification (with suitable decision regions) is equivalent to the energy detector from~\eqref{eq:ml_ed}.
        Similarly, for the high SNR detector in~\eqref{eq:ml_hsnr}, its corresponding quadratic statistic is
        \begin{equation}
            \hspace*{-0.5em}\mathbf{A}_{\mathrm{HSNR}}\triangleq\frac{\mathbf{\Gamma}^{-1}}{N}\implies\widehat{\varepsilon}_{\mathrm{HSNR}}(\mathbf{r})=\frac{\herm{\mathbf{r}}\mathbf{\Gamma}^{-1}\mathbf{r}-\trace{\mathbf{\Gamma}^{-1}}{false}}{N}.
        \end{equation}
        Notice that the affine term in both cases has been set to $c=1-\trace{\mathbf{A}\overline{\mathbf{C}}}{false}$ to more easily compare them with the \acrshort{qmmse}.
        Remarkably, this makes them conditionally unbiased.

        An interesting aspect regarding $\widehat{\varepsilon}_{\mathrm{QMMSE}}$ is that, in the high \acrshort{snr} regime, its quadratic term matrix becomes
        \begin{equation}
            \lim_{\alpha\to\infty}\mathbf{A}_{\mathrm{QMMSE}}=\lim_{\alpha\to\infty}\frac{1}{1+\frac{1}{\sigma_{\varepsilon}^2}+N}\mathbf{\Gamma}^{-1}.
        \end{equation}
        This is a scaled version of $\mathbf{A}_{\mathrm{HSNR}}$ and, in fact, both matrices coincide for asymptotically large $N$.
        Therefore, at high \acrshort{snr}, both $\widehat{\varepsilon}_{\mathrm{QMMSE}}$ and $\widehat{\varepsilon}_{\mathrm{HSNR}}$ will present the same performance and error floor.

        Another important property of $\widehat{\varepsilon}_{\mathrm{ED}}$, $\widehat{\varepsilon}_{\mathrm{HSNR}}$ and $\widehat{\varepsilon}_{\mathrm{BQUE}}$ is that they all converge to the same estimator when the channel is assumed to be isotropic:
        \begin{equation}
            \widehat{\varepsilon}(\mathbf{r})=\frac{1}{\alpha}\left(\frac{1}{N}\|\mathbf{r}\|^2-1\right).\label{eq:isotropic}
        \end{equation}
        Similarly, $\widehat{\varepsilon}_{\mathrm{QMMSE}}$ becomes
        \begin{equation}
            \widehat{\varepsilon}'(\mathbf{r})=\frac{\|\mathbf{r}\|^2+\frac{(\alpha+1)^2}{\sigma_{\varepsilon}^2\alpha}+\alpha-N}{\frac{(\alpha+1)^2}{\sigma_{\varepsilon}^2\alpha}+\alpha+N\alpha},
        \end{equation}
        which is a scaled and translated version of~\eqref{eq:isotropic}, \ie they define equivalent detectors with appropriate decision regions.

\section{Symbol detection} \label{sec:detection}
    Once the energy estimation phase has been studied, we now move on to the second step in the architecture proposed in Section~\ref{sec:estimation}: symbol detection.
    For a given constellation, we describe the detection regions considering that the energy statistic asymptotically follows a Gaussian distribution.
    Afterwards, we analyze the probability of error of quadratic detectors and present a decision-directed design that leverages the \acrshort{bque} to provide an improved performance.

    \subsection{Detection regions} \label{sec:detection_regions}
        Quadratic energy estimators of the form~\eqref{eq:full_quad} can be expressed as a summation:
        \begin{equation} \label{eq:estimator_structure}
            \widehat{\varepsilon}\left(\mathbf{r}\right) = \herm{\mathbf{r}} \mathbf{A} \mathbf{r} + c = \sum\limits_{n=1}^{N} a_{n} \left|r_n\right|^2 + c_n,
        \end{equation}
        where $a_{n} \triangleq [\mathbf{A}]_{n,n}$ and $\sum_nc_n\triangleq c$.
        Particularizing for the estimators presented in the previous section, we can set $c\triangleq1-\trace{\mathbf{A}(\mathbf{\Gamma}+\id{N})}{false}$.
        Notice how \eqref{eq:estimator_structure} produces a real output.
        Therefore, defining a decision region $\mathcal{R}_x$ for each $x\in\mathcal{X}$ consists in determining detection thresholds on the real line:
        \begin{equation} \label{eq:detection_problem}
        \widehat{x} =
            \begin{cases}
                x_1,\quad &\widehat{\varepsilon} \leq \tau_1,\\
                x_i,\quad &\tau_{i-1} < \widehat{\varepsilon} < \tau_i,\\
                x_{M},\quad &\tau_{M-1} \leq \widehat{\varepsilon}.
            \end{cases}
        \end{equation}
        The thresholds that minimize the error probability, $\mathcal{T}\triangleq\{\tau_i\}_{1\leq i\leq M-1}$, are found at the intersection between the densities of the estimator output conditioned to every transmitted symbol of $\mathcal{X}$.
        
        Since $\widehat{\varepsilon}(\bsf{r}|\varepsilon)$ is a quadratic form of a complex normal vector, it has a \textit{generalized chi-squared} distribution~\cite{mathai_quadratic_1992}.
        Its \acrshort{pdf} can be obtained analytically in specific cases~\cite{Hammarwall2008,Bjornson2009}.
        Nevertheless, the resulting expressions for general quadratic detectors and arbitrary correlation models are usually very involved~\cite[Ch.~1]{cox_renewal_1982}, \cite{brehler_asymptotic_2001} and do not allow for a simple derivation of detection thresholds.

        In the context of massive \acrshort{simo}, we can exploit the asymptotic properties of $\widehat{\varepsilon}(\bsf{r}|\varepsilon)$ for large $N$.
        Invoking the \acrfull{clt},
        \begin{equation}
            \widehat{\varepsilon}\left(\bsf{r}|\varepsilon\right) \xrightarrow{\mathrm{d}} \mathcal{N}\left(1-(1-\varepsilon)\trace{\mathbf{A}\mathbf{\Gamma}}{true},\, \frob{\mathbf{A}\left(\varepsilon\mathbf{\Gamma}+\mathbf{I}_{N}\right)}{true}^2\right),
        \end{equation}
        as $N \to \infty$.
        For the unbiased estimators, $\trace{\mathbf{A}\mathbf{\Gamma}}{true}=1$, thus the Gaussian density function is centered at the symbol energy level.
        Observe that the terms in sum \eqref{eq:estimator_structure} are independent but not identically distributed.
        To ensure the \acrshort{clt} can be applied in this scenario, it is sufficient to prove that \textit{Lyapunov's condition} is satisfied.
        In Appendix~\ref{ap:lyapunov} we show that it does indeed hold, thus validating the use of the \acrshort{clt}.
        This result motivates employing a Gaussian approximation for the densities of $\widehat{\varepsilon}(\bsf{r}|\varepsilon)$ which, on its turn, implies that the bound in Eq.~\eqref{eq:mi_bound} is asymptotically an equality.

        In the Gaussian case, finding the intersection points reduces to finding the roots of a second-degree polynomial for each pair of adjacent densities.
        In general, each polynomial has two roots, but we are only interested in the one between the two likelihoods.
        Taking into account that the variance increases when the symbol energy does as well,
        we conclude that the desired root is the greatest one, thus the maximum of each pair must be taken to find the thresholds.
        This procedure is summarized in Algorithm~\ref{alg:gaussian_thresholds}.
        \begin{algorithm}[H]
             \caption{Thresholds between Gaussian Likelihoods}
             \label{alg:gaussian_thresholds}
             \begin{algorithmic}[1]
                \Require constellation $\mathcal{X}$, matrix $\mathbf{A}$, spectrum $\mathbf{\Gamma}$
                \Ensure  thresholds $\mathcal{T} = \left\{\tau_1, \dots, \tau_{M-1}\right\}$
                \Procedure{NormalIntersection}{$\mu_1,\mu_2,\sigma^2_1,\sigma^2_2$}
                    \State $a = 1/\sigma^2_2 - 1/\sigma^2_1$
                    \State $b = 2 \cdot (\mu_1/\sigma^2_1 - \mu_2/\sigma^2_2)$
                    \State $c = \mu_2^2 / \sigma^2_2 - \mu_1^2/\sigma^2_1 + \log(\sigma^2_2/\sigma^2_1)$
                    \State \textbf{return} $\mathrm{roots}(\mathcal{P}(c,b,a))$
                \EndProcedure
                \For{$i = 1:M-1$}
                    \State $\mu_1 = 1-\left(1-\varepsilon_i\right)\trace{\mathbf{A\Gamma}}{true}$
                    \State $\mu_2 = 1-\left(1-\varepsilon_{i+1}\right)\trace{\mathbf{A\Gamma}}{true}$
                    \State $\sigma^2_1 = \trace{\mathbf{A}^2(\varepsilon_i\mathbf{\Gamma}+\mathbf{I}_{N})^2}{true}$
                    \State $\sigma^2_2 = \trace{\mathbf{A}^2(\varepsilon_{i+1}\mathbf{\Gamma}+\mathbf{I}_{N})^2}{true}$
                    \State $\tau_i = \max(\Call{NormalIntersection}{\mu_1,\mu_2,\sigma^2_1,\sigma^2_2})$
                \EndFor
                \State \textbf{return} $\mathcal{T}$
            \end{algorithmic} 
        \end{algorithm}        

    \subsection{Probability of detection error} \label{sec:prob_error_optimal}
        From \eqref{eq:detection_problem} we observe that the error probability can be computed as:
        \begin{multline}
            \mathrm{P}_{\epsilon} = \frac{1}{M} \bigg{(}\prob{\widehat{\varepsilon}(\bsf{r}|\varepsilon_1) > \tau_1}{true} + \prob{\widehat{\varepsilon}(\bsf{r}|\varepsilon_M) < \tau_{M-1}}{true}\\
            + \sum_{i=2}^{M-1} \prob{\widehat{\varepsilon}(\bsf{r}|\varepsilon_i) < \tau_{i-1}}{true} + \prob{\widehat{\varepsilon}(\bsf{r}|\varepsilon_i) > \tau_{i}}{true}\bigg{)}.
        \end{multline}
        Under the Gaussian assumption, each tail probability is approximated with the $\mathrm{Q}$-function:
        \begin{align}
            &\prob{\widehat{\varepsilon}(\bsf{r}|\varepsilon_i)<\tau_{i-1}}{true}\approx\mathrm{Q}\left(\frac{1-(1-\varepsilon_i)\trace{\mathbf{A}\mathbf{\Gamma}}{true}-\tau_{i-1}}{\frob{\mathbf{A}\left(\varepsilon_i\boldsymbol{\Gamma}+\id{N}\right)}{true}}\right)\nonumber\\
            &\prob{\widehat{\varepsilon}(\bsf{r}|\varepsilon_i)>\tau_{i}}{true}\approx\mathrm{Q}\left(\frac{\tau_{i}-1+(1-\varepsilon_i)\trace{\mathbf{A}\mathbf{\Gamma}}{true}}{\frob{\mathbf{A}\left(\varepsilon_i\boldsymbol{\Gamma}+\id{N}\right)}{true}}\right),\label{eq:q_func}
        \end{align}
        with $\varepsilon_i$ defined as in \eqref{eq:constellation}.
        Thus, the error probability is approximately given by:
        \begin{multline}
                \mathrm{P}_{\epsilon}\approx\frac{1}{M}\Bigg{(}\sum_{i=2}^{M}\mathrm{Q}\bigg{(}\frac{1-(1-\varepsilon_i)\trace{\mathbf{A}\mathbf{\Gamma}}{true}-\tau_{i-1}}{\frob{\mathbf{A}\left(\varepsilon_i\boldsymbol{\Gamma}+\id{N}\right)}{true}}\bigg{)}\\
                +\sum_{j=1}^{M-1}\mathrm{Q}\bigg{(}\frac{\tau_{j}-1+(1-\varepsilon_j)\trace{\mathbf{A}\mathbf{\Gamma}}{true}}{\frob{\mathbf{A}\left(\varepsilon_j\boldsymbol{\Gamma}+\id{N}\right)}{true}}\bigg{)}\Bigg{)}.
            \label{eq:error_probability}
        \end{multline}
        In expression~\eqref{eq:error_probability} we observe that the error probability increases with the norms $\frob{\mathbf{A}(\varepsilon_j\boldsymbol{\Gamma}+\id{N})}{false}$,
        which correspond to the \acrshort{mse} of the \acrshort{bque} for each symbol.
        This dependence shows that the information-theoretic criteria proposed in~\ref{sec:information-criteria} is indeed justified.
        Nevertheless, optimizing \acrshort{mse} on average (\ie as in \acrshort{qmmse}) is not consistent with the fact that the \acrshort{pep} of each symbol has a different effect on the error probability for each \acrshort{snr} level.
        
    \subsection{Assisted BQUE (ABQUE)}
        In order to overcome the previous issue, we now propose to assist the \acrshort{bque} by replacing the true transmitted symbol known by the genie-aided decoder by the final decision delivered by the \acrshort{ed}.
        By doing so, we build a hard-decision detector which exhibits some computational advantages compared with its soft alternative (plugging the energy estimation without deciding).
        On the one hand, assisting with hard decisions is prone to error boosting at low-\acrshort{snr}.
        However, the simulations in Section~\ref{sec:numerical_results} will show that this effect is negligible except for extremely high antenna correlation.
        On the other hand, in the hard-decision scheme, the symbol plugged into the \acrshort{bque} belongs to a known constellation, thus matrices $\mathbf{A}_{\mathrm{BQUE},\,i}$ for each symbol only need to be computed once and can be stored for later uses:
        \begin{equation}
            \mathbf{A}_{\mathrm{BQUE},\,i} = \frac{\mathbf{\Gamma}\cov{\bsf{r}|\varepsilon_i}^{-2}}{\frob{\mathbf{\Gamma}\cov{\bsf{r}|\varepsilon_i}^{-1}}{false}^2}, \quad 1 \leq i \leq M.
        \end{equation}
        This result has further implications.
        Given that the thresholds of the proposed quadratic detectors only depend on the constellation and $\mathbf{A}$, if $\mathbf{A}$ can be computed only once, thresholds can also be computed only once.
        
        The decision-directed nature of the \acrshort{abque} makes its theoretical analysis much more involved than that of the quadratic detectors considered in previous sections.
        Nonetheless, the \acrshort{bque} can serve as a reliable benchmark to assess it, by providing an analytic bound on its error probability.
        Simulations in the next section show that both detectors perform very similarly (in terms of \acrshort{ser}) in a variety of scenarios.

\section{Numerical results} \label{sec:numerical_results}
    In this section, we provide some simulations with two different goals.
    First, to illustrate performance aspects that are not apparent from analytic results presented in the preceding sections;
    in particular, we display how the \textit{outage probability} performance of statistical \acrshort{csi}-aware detectors benefits from channel hardening.
    Second, to numerically validate the theoretical results presented in the previous sections and compare the various detectors in terms of average \acrshort{ser} performance.
    
    In all simulations we have employed signal \eqref{eq:rx_signal} assuming the exponential correlation channel model described in~\cite{han_constellation_2022}:
    \begin{equation}
        [\cov{\bsf{h}}]_{k,l} = c_{k,l} = \begin{cases}
            \rho^{l-k},\quad k \leq l,\\
            c_{l,k}^*,\quad k > l,
        \end{cases}
    \end{equation}
    with correlation coefficient $\rho=0.7$ (unless stated otherwise).
    Symbols from a uniform unipolar 8-\acrshort{ask} constellation have been transmitted with average power equal to one.
    We have chosen a standard constellation instead of one optimized to the channel statistics (\eg see~\cite{han_constellation_2022}) to better portray a realistic scenario, with a low complexity transmitter that is unaware of \acrshort{csi}.
    It has the added benefit of being robust to \acrshort{snr} estimation errors in transmission.

    \subsection{Outage probability}
        \begin{figure}[tb]
            \centering
            \includegraphics[width=\columnwidth]{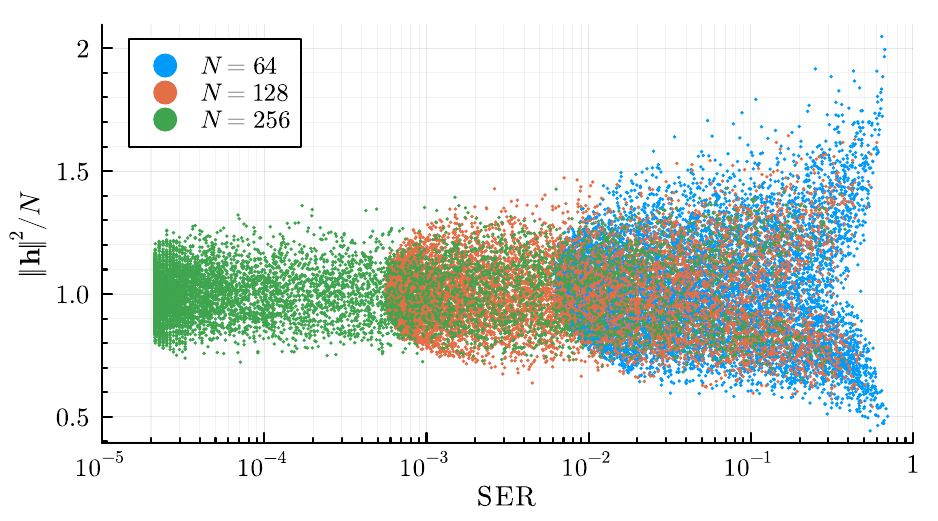}
            \caption{Scatter plot to assess the relevance of the channel norm as an indicator of \acrshort{ser} performance. The \acrshort{bque} error probability for $10^4$ different channels at $\mathrm{SNR} = \SI{10}{\decibel}$ is depicted.}
            \label{fig:ser_channel}
        \end{figure}
        \begin{figure}[tb]
            \centering
            \includegraphics[width=\columnwidth]{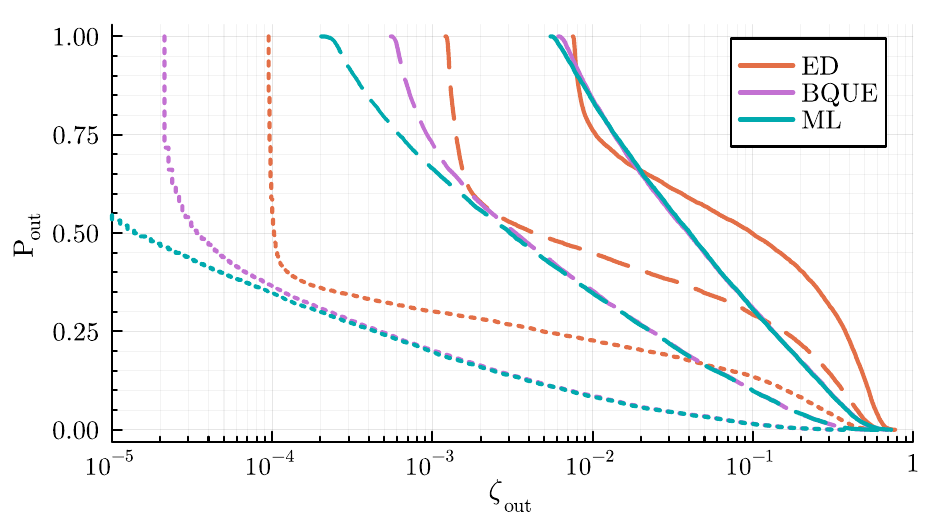}
            \caption{Outage probability for a \acrshort{ser} threshold at $\mathrm{SNR} = \SI{10}{\decibel}$ for $N=64$ (solid lines), $N=128$ (dashed lined) and $N=256$ (dotted lines).}
            \label{fig:ser_outage}
        \end{figure}
        
        An outage event occurs when the instantaneous \acrshort{ser} of a system is above a certain threshold $\zeta_{\mathrm{out}}$, given a channel realization~\cite{Lombardo1999}.
        In the context of \acrshort{urllc}, the outage probability is a relevant quality-of-service metric because it is intimately related to the reliability and latency of a communication setup~\cite{Popovski2019}.
        Within the one-shot scheme considered in this paper, we define it as follows:
        \begin{equation}
            \mathrm{P}_{\mathrm{out}}\triangleq\int_{\mathcal{O}}f_{\bsf{h}}(\mathbf{h})\der\mathbf{h},
        \end{equation}
        where $\mathcal{O} = \{\mathrm{P}_{\epsilon}(\mathbf{h})>\zeta_{\mathrm{out}} \, | \, \mathbf{h} \in \mathbb{C}^{N}\}$.
        In coherent communications, it is equivalent to the probability that the instantaneous \acrshort{snr} drops below a certain value~\cite[Eq.~(6.46)]{goldsmith_wireless_2005}.
        On the contrary, this correspondence does not hold in noncoherent systems of the kind considered herein, as it is clear from Fig.~\ref{fig:ser_channel}.
        We can observe that the highest \acrshort{ser} values correspond to channel realizations that deviate the most from the expected value, from both above and below.
        Channel hardening counteracts this deviation from the mean: the higher the number of antennas at the receiver, the more concentrated the channel realizations are (in the norm), and the lower their associated \acrshort{ser} values are.
        
        Fig.~\ref{fig:ser_outage} illustrates how this hardening affects outage probability.
        In general terms, employing more antennas stabilizes the channel statistics and reduces the chances of dealing with an outage event.
        Although \acrshort{ed} benefits from this property, its hardening gains are less pronounced than \acrshort{ml} and \acrshort{bque} ones.
        These latter two share similar performance unless the quality requirement becomes very stringent, in which case \acrshort{ml} outperforms \acrshort{bque}.
        It is worth to remark that the minimum \acrshort{ser} in Fig.~\ref{fig:ser_channel} coincides with the threshold $\zeta_{\mathrm{out}}$ for which the \acrshort{bque} outage probability reaches its maximum in Fig.~\ref{fig:ser_outage}.
        
        Another relevant phenomenon that can be observed in Fig.~\ref{fig:ser_outage} is the fact there exist channel realizations for which the \acrshort{ed} outperforms the \acrshort{ml} detector.
        Indeed, (unconditional) \acrshort{ml} detection is optimal in terms of average error probability, but not necessarily when conditioned to a specific channel realization.

    \subsection{SER analysis}
        In order to evaluate the validity of the error probability expression from~\eqref{eq:error_probability}, it is compared against Monte Carlo results of the high \acrshort{snr}, \acrshort{ed}, \acrshort{qmmse} and \acrshort{bque} detectors.
        Analytical expressions are drawn as continuous lines, whereas simulation results are painted as dotted lines with markers sharing the same color.
        \acrshort{ml} and \acrshort{abque} detectors are also assessed numerically.
        
\begin{figure}[tb]
            \centering
            \includegraphics[width=\columnwidth]{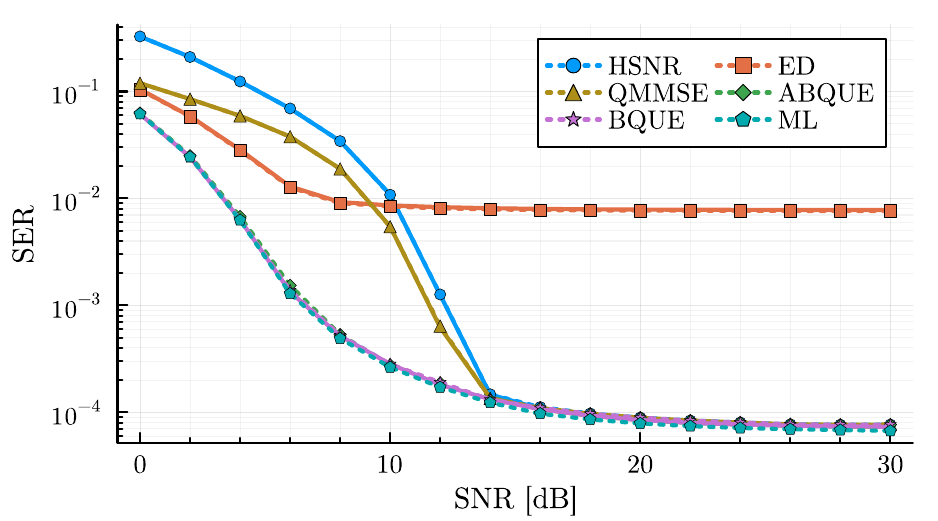}
            \caption{\acrshort{ser} of the presented detectors in terms of \acrshort{snr} for $N=512$.}
            \label{fig:ser_snr}
        \end{figure}

        In Fig.~\ref{fig:ser_snr}, the performance of all detectors discussed previously is depicted as a function of the \acrshort{snr} for $N=512$ antennas.
        We have computed the thresholds for each quadratic detector and then classified the estimated powers,
        as described in Section~\ref{sec:detection_regions}.
        A first observation of these results reveals that the \acrshort{ed} error floor is much higher than that of the other quadratic detectors, which share it with the genie-aided detector.
        This error floor is very close to the one obtained with \acrshort{ml} detection.
        The reason behind the slight discrepancy is that, although the \acrshort{clt} holds, the Gaussian likelihoods assumed in Algorithm~\ref{alg:gaussian_thresholds} present inaccuracies that affect the threshold positioning, thus resulting in an increased error floor.
        Another notable outcome is that the \acrshort{abque} exhibits a performance very close to that of the oracle (\ie \acrshort{bque}) with just a narrow increase in complexity.
    
        \begin{figure}[tb]
            \centering
            \includegraphics[width=\columnwidth]{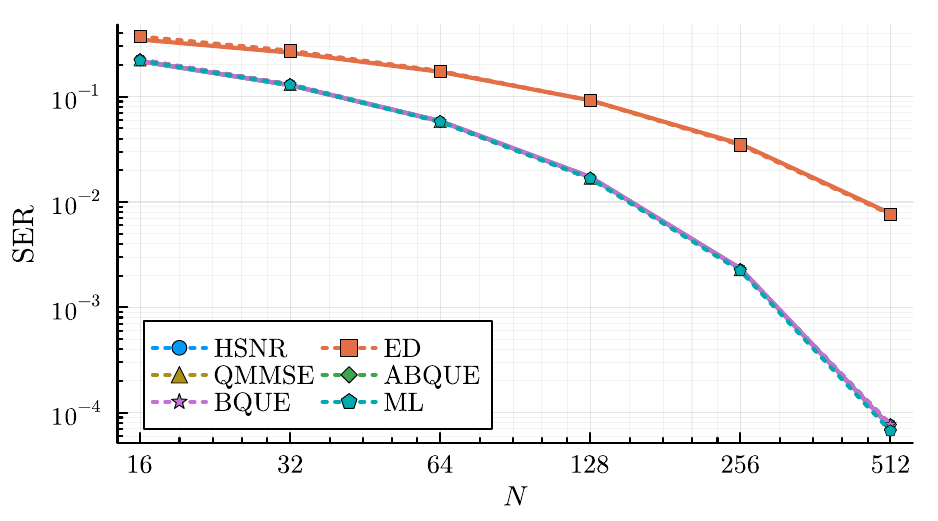}
            \caption{Floor level (\ie \acrshort{ser} at $\mathrm{SNR}=\SI{30}{\decibel}$) of the presented detectors in terms of $N$.}
            \label{fig:ser_n}
        \end{figure}
        \begin{figure}[tb]
            \centering
            \includegraphics[width=\columnwidth]{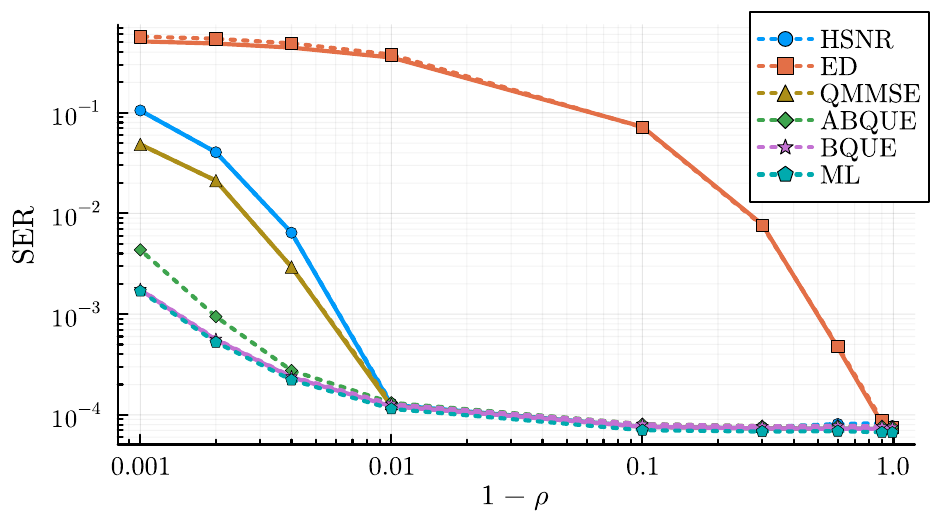}
            \caption{Floor level (\ie \acrshort{ser} at $\mathrm{SNR}=\SI{30}{\decibel}$) of the presented detectors in terms of channel correlation, for $N=512$. The plot is with respect to $1-\rho$ for better visualization.}
            \label{fig:ser_r}
        \end{figure}
        
        Figs.~\ref{fig:ser_n} and~\ref{fig:ser_r} illustrate the error floor level by depicting the \acrshort{ser} of the different detectors at $\mathrm{SNR}=\SI{30}{\decibel}$ in terms of the number of antennas and the correlation coefficient, respectively.
        In the former, we can observe the penalization suffered by \acrshort{ed}.
        By being agnostic to channel statistics, it is not able to fully exploit the increasing number of antennas.
        As a result, its error floor is higher than the other methods' and it decreases at a slower rate, as well.
        The other quadratic detectors and \acrshort{abque} all display error levels similar to \acrshort{ml}.
        Regarding the \acrshort{ser} behavior in terms of channel correlation, the statistical \acrshort{csi}-aware detectors display remarkable robustness, since their error floors do not change significantly for $0\leq\rho\leq0.9$.
        Moreover, this \acrshort{ser} is shared with the \acrshort{ml} detector for $\rho < 0.99$ (up to the gap mentioned above).
        On the other hand, the error probability of the \acrshort{ed} is acutely sensitive to channel correlation.
        For instance, at $\rho = 0.7$, its performance has degraded by almost a factor of 100, compared with the uncorrelated case.
        
        A final note on these results is the remarkable accuracy of the approximated error probability expressions derived from~\eqref{eq:error_probability}.
        Indeed, in Fig.~\ref{fig:ser_n} we can observe that the difference between the theoretical and the numerical \acrshort{ser} is only barely noticeable for small $N$, illustrating how the Gaussian approximation improves when the number of antennas increases.

\section{Comparison of detection schemes}\label{sec:comparison}
    In this section, we undertake a comprehensive comparison between the established \acrshort{ml} and energy detectors and the novel \acrshort{bque}, \acrshort{qmmse} and \acrshort{abque} detectors introduced in this study. To facilitate this comparison, Table~\ref{tab:detectors_comparison} outlines six distinct properties for each detector:
    \begin{enumerate}
        \item \textbf{Complexity:} refers to the computational complexity of the entire detection process.
        \item \textbf{Performance:} denotes the efficiency of the detector in terms of error probability.
        \item \textbf{Graceful degradation:} reflects the resilience of the detector in front of small changes in channel correlation.
        \item \textbf{Tractable:} indicates the possibility of mathematically analyzing the detector and its error probability.
        \item \textbf{Implementable:} describes whether the detector depends on the transmitted symbol or not.
        \item \textbf{Agnostic to \acrshort{csir}:} specifies whether the detector needs to know the channel covariance.
    \end{enumerate}

    It is worth to mention that no single detector emerges as superior across all areas.
    Instead, there exists a trade-off among the aforementioned properties.
    For instance, while the \acrshort{ed} has a very low complexity compared to other detectors, its performance is severely degraded for even slightly correlated channels (see Fig.~\ref{fig:ser_r}).
    
    Conversely, both the \acrshort{ml} and \acrshort{bque} detectors exhibit comparable performance and complexity.
    However, only the latter has an analytic expression for the error probability (see Sec.~\ref{sec:prob_error_optimal}).
    Furthermore, the tractability of \acrshort{bque} can be extended to \acrshort{abque} when the channel correlation is not extremely high (see Fig.~\ref{fig:ser_r}).

    Finally, the \acrshort{qmmse} detector behaves like \acrshort{bque} at moderate and high \acrshort{snr}, but its performance is inferior to that of the \acrshort{ed} at low \acrshort{snr}.
    Even so, its resilience in the face of high channel correlation must not be overstated.
    A viable implementation that exhibits a high performance and is tractable is to leverage the \acrshort{ed} at low \acrshort{snr} and switch to the \acrshort{qmmse} at moderate and high regimes.

    \begin{table*}
        \renewcommand{\arraystretch}{1.3}
        \centering
        \caption{Summary of detection schemes properties.}
        \label{tab:detectors_comparison}
        \begin{tabular}{c|cccccc}
            Detector & Complexity & Performance & \begin{tabular}[c]{@{}c@{}}Graceful\\ degradation\end{tabular} & Tractable & Implementable & \begin{tabular}[c]{@{}c@{}}Agnostic to\\ \acrshort{csir}\end{tabular} \\ \hline\hline
            \acrshort{ml} & $O(N^2) + O(MN)$ & Optimal & Yes & No & Yes & No\\ \hline
            \acrshort{ed} & $O(N) + O(M)$ & \begin{tabular}[c]{@{}c@{}}High loss\\[-0.5ex] $(\rho \neq 0)$\end{tabular} & No & Yes & Yes & Yes\\ \hline
            \acrshort{bque} & $O(N^2) + O(N) + O(M)$ & Negligible loss & Yes & Yes & No & No\\ \hline
            \acrshort{qmmse} & $O(N^2) + O(N) + O(M)$ & \begin{tabular}[c]{@{}c@{}}Negligible loss\\[-0.5ex] (moderate \& high \acrshort{snr})\end{tabular} & \begin{tabular}[c]{@{}c@{}}Yes\\[-0.5ex] $(\rho < \num{0.99})$\end{tabular} & Yes & Yes & No\\ \hline
            \acrshort{abque} & $O(N^2) + O(N) + O(M)$ & Negligible loss & Yes & No & Yes & No
        \end{tabular}
    \end{table*}

\section{Conclusions} \label{sec:conclusions}
    In this paper, we have analyzed the fundamental limitations of one-shot \acrshort{simo} noncoherent communication when an arbitrary \acrshort{pam} constellation is considered.
    We have also introduced a quadratic framework that generalizes the \acrshort{ed} commonly used in the literature.
    We have derived an analytic approximation for the error probability of any detector exhibiting the quadratic structure proposed, as a sum of $\mathrm{Q}$-functions.
    An improved scheme based on the combination of quadratic detectors has also been presented.
    Their performance in terms of average \acrshort{ser} and outage probability has been tested through several Monte Carlo experiments, as well as the validity of the \acrshort{ser} approximations.

    We can outline some future research lines that arise from this work.
    The most straightforward one is to design \acrshort{pam} constellations optimized for a particular quadratic detector, by means of minimizing the error probability approximation~\eqref{eq:error_probability}.
    Similar approaches are studied in~\cite{Manolakos2016,gao_energy-efficient_2019,han_constellation_2022}.
    Other potential extensions of the presented analysis include the design of codes across multiple channel uses and suitable detection schemes~\cite{Knott2015,chowdhury_scaling_2016}.
    Finally, a possible application of the proposed framework is in noncoherent energy detection of index modulations~\cite{Fazeli2020, Fazeli2022}.
    For instance, considering a frequency selective channel and a multicarrier modulation such as \acrfull{ofdm}, information can be conveyed by assigning different transmission power levels to different sets of subcarriers.

\appendix
    \subsection{Proof of Theorem 1}\label{ap:thm1}
        The \acrshort{pep} can be upper bounded making use of \textit{Cantelli's inequality}~\cite{ngo_joint_2022}:
        \begin{equation}
            \mathrm{P}_{a\to b}\leq\left(1+\Delta_{a,b}\right)^{-1},
        \end{equation}
        where
        \begin{equation}
            \Delta_{a,b}\triangleq\frac{\expec{\mathrm{L}_{a,b}\left(\bsf{r}|x_a\right)}{\bsf{r}|x_a}{true}^2}{\var{\mathrm{L}_{a,b}\left(\bsf{r}|x_a\right)}{\bsf{r}|x_a}{true}}.\label{eq:delta}
        \end{equation}
        We must check whether this term grows without bounds, so that the \acrshort{pep} vanishes.
        This is intimately related to the channel hardening phenomenon exhibited by large arrays; notice how~\eqref{eq:delta} is analogous to~\cite[Eq.~(10)]{chen_channel_2018}.
        
        The elements involved in the computation of~\eqref{eq:delta} are:
        \begin{equation}
            \begin{aligned}
                &\expec{\mathrm{L}_{a,b}\left(\bsf{r}|x_a\right)}{\bsf{r}|x_a}{true}=\sum_{n=1}^{N}\left(\lambda_n-1-\log\lambda_n\right),\\
                &\var{\mathrm{L}_{a,b}\left(\bsf{r}|x_a\right)}{\bsf{r}|x_a}{true}=\sum_{n=1}^{N}\left(\lambda_n-1\right)^2,
            \end{aligned}\label{eq:expected}
        \end{equation}
        where
        \begin{equation}
            \lambda_n\triangleq\frac{|x_a|^2\gamma_n+1}{|x_b|^2\gamma_n+1}.\label{eq:lambda}
        \end{equation}
        Thus, we are left with
        \begin{equation}
            \Delta_{a,b}=\frac{\left(\sum_{n=1}^{N}\left(\lambda_n-1-\log\lambda_n\right)\right)^2}{\sum_{n=1}^{N}\left(\lambda_n-1\right)^2}\triangleq\frac{u_N}{d_N}.\label{eq:deflect}
        \end{equation}
        
        The proof is based on the \textit{Stolz--Cesàro theorem}~\cite[Th.~2.7.2]{choudary_real_2014}, so various requirements must be fulfilled to apply it.
        Both $\{u_n\}_{n\geq1}$ and $\{d_n\}_{n\geq1}$ are sequences of real numbers.
        In addition, $\{d_n\}_{n\geq1}$ must be strictly monotone and divergent, which is guaranteed due to the \textit{test for divergence} (\ie the contrapositive of~\cite[Lemma~4.1.2]{choudary_real_2014}), by ensuring
        \begin{equation}
            (\lambda_n-1)^2>0,\quad\forall n\in\{1,\dots,\mathrm{\Theta}(N)\},
        \end{equation}
        such that $\lim_{N\to\infty}\mathrm{\Theta}(N)=\infty$.
        This condition reduces to
        \begin{equation}
            \lambda_{n}\neq1\iff|x_a|^2\gamma_n+1\neq|x_b|^2\gamma_n+1.
        \end{equation}
        Thus, under the premise of the spectrum $\{\gamma_n\}_{n\geq1}$ being nonnegative and bounded, our only requirement is
        \begin{equation}
            |x_a|^2\neq|x_b|^2\iff x_a\neq x_b,\quad x_a,x_b\in\mathcal{X},\label{eq:ufc}
        \end{equation}
        which is consistent with the loss of phase information observed in the \acrshort{ml} detector \eqref{eq:ml_det}.
        
        Property~\eqref{eq:ufc} is a specific instance of a broader concept known as \textit{unique identification}.
        It has been recognized throughout the literature under various forms~\cite{Dong2017, li_design_2019, Han2021}\footnote{These works use terms like \textit{unique factorization}, \textit{unique identification} or \textit{unique determination} to define similar ideas, while others employ them to refer to different concepts~\cite{Zhang2011, Xiong2012, Xia2013}.}, but we abide by the one presented in~\cite[Prop.~1]{Dong2023}.
        In general terms, a constellation is uniquely identifiable if there is a one-to-one correspondence between each of its symbols and a distinct second-order statistical structure at the receiver.
        Indeed, particularized for our model~\eqref{eq:rx_signal} amounts to
        \begin{equation}
            x_a\cov{\bsf{h}}x_a^*\neq x_b\cov{\bsf{h}}x_b^*\iff x_a\neq x_b,\quad x_a,x_b\in\mathcal{X},
        \end{equation}
        which is equivalent to~\eqref{eq:ufc}.
        Not fulfilling it would collapse the \acrshort{llr} in~\eqref{eq:llr} and thus positively lower bound the \acrshort{pep}: it is a \textit{necessary condition} for (asymptotically) error-free \acrshort{ml} detection.
        As it will be clear by the end of this proof, it is also a \textit{sufficient condition} to achieve vanishing error probability as the number of receiving antennas increases.

        Having dealt with the behavior of $\{d_n\}_{n\geq1}$, we are now in position to apply the Stolz--Cesàro theorem:
        \begin{multline}\label{eq:stolz-cesaro}
            \lim_{N\to\infty}\Delta_{a,b} = \lim_{N\to\infty}\frac{u_{N+1}-u_N}{d_{N+1}-d_N}\\
            =\lim_{N\to\infty}\frac{\eta\left(\lambda_{N+1}\right)\left(\eta\left(\lambda_{N+1}\right)+2\sum_{n=1}^N\eta\left(\lambda_n\right)\right)}{\left(\lambda_{N+1}-1\right)^2},
        \end{multline}
        where $\eta(x)\triangleq x-1-\log x$ is a nonnegative convex function that is 0 only at $x=1$.
        Each term involved in~\eqref{eq:stolz-cesaro} is positive by property~\eqref{eq:ufc}, and bounded due to the boundedness of $\{\gamma_n\}_{n\geq1}$.
        Then, the previous limit simplifies to
        \begin{equation}
            \lim_{N\to\infty}\Delta_{a,b}=\lim_{N\to\infty}\frac{2\eta\left(\lambda_{N+1}\right)}{\left(\lambda_{N+1}-1\right)^2}\sum\limits_{n=1}^N\eta\left(\lambda_{n}\right),
        \end{equation}
        which will grow without bounds since the test for divergence holds for the sum involved.
        This completes the proof.\hfill\IEEEQED

    \subsection{Proof of Theorem 2}\label{ap:thm2}
        Referring to \eqref{eq:pep}, the \acrshort{pep} can be expressed as~\cite[Sec.~2.4]{levy_principles_2008}
        \begin{equation}
            \mathrm{P}_{a\to b}=\int_{\mathcal{D}}f_{\bsf{r}|x_a}(\mathbf{r})\der\mathbf{r},\label{eq:pep_int}
        \end{equation}
        where $\mathcal{D}\triangleq\left\{\mathbf{r}\in\mathbb{C}^N:f_{\bsf{r}|x_b}(\mathbf{r})\geq f_{\bsf{r}|x_a}(\mathbf{r})\right\}$.
        With some simple manipulations, we can see that the boundary of this region defines an $(N-1)$-dimensional complex ellipsoid:
        \begin{equation}
            \partial\mathcal{D}=\left\{\mathbf{r}\in\mathbb{C}^N:\herm{\mathbf{r}}\mathbf{K}\mathbf{r}=1\right\},\,\,\mathbf{K}\triangleq\frac{\cov{\bsf{r}|x_b}^{-1}-\cov{\bsf{r}|x_a}^{-1}}{\log\bigl|\cov{\bsf{r}|x_a}\cov{\bsf{r}|x_b}^{-1}\bigr|}.\label{eq:ellipsoid}
        \end{equation}
        Without loss of generality\footnote{If the channel is rank deficient, an analogous analysis can be performed on a lower dimensional subspace.}, we assume the channel is full-rank (\ie $\gamma_n>0$ for $n=1,\dots,N$).
        If condition \eqref{eq:puf} is fulfilled, $\mathbf{K}$ is always positive definite, ensuring the ellipsoid exists.

        When $|x_a|^2>|x_b|^2$, $\mathcal{D}$ is the closure of the set of points inside ellipsoid~\eqref{eq:ellipsoid}.
        With a change of variable $\mathbf{r}\triangleq|\mathbf{K}|^{\frac{1}{2N}}\mathbf{K}^{-\frac{1}{2}}\mathbf{s}$, we can map it to a $N$-dimensional closed ball with the same volume:
        \begin{equation}
            \mathcal{D}\mapsto\mathcal{U}\triangleq\left\{\mathbf{s}\in\mathbb{C}^N:\|\mathbf{s}\|^2\leq|\mathbf{K}|^{-\frac{1}{N}}\right\}.\label{eq:change}
        \end{equation}
        Then, the \acrshort{pep} becomes:
        \begin{equation}
            \mathrm{P}_{a\to b}=\int_{\mathcal{U}}\frac{\exp\left(-\herm{\mathbf{s}}\mathbf{\Omega}^{-1}\mathbf{s}\right)}{\pi^{N}|\cov{\bsf{r}|x_a}|}\der\mathbf{s},\quad\mathbf{\Omega}\triangleq\frac{\cov{\bsf{r}|x_a}\mathbf{K}}{|\mathbf{K}|^{\frac{1}{N}}},
        \end{equation}
        which is the integral of a multivariate complex Gaussian function.
        We can lower bound it with the integral in $\mathcal{U}$ of a narrower Gaussian function that is tangent to it along the direction with the lowest eigenvalue of $\mathbf{\Omega}$, named $\underline{\omega}$.
        It is proportional to the \acrshort{pdf} of $\bsf{t}\sim\mathcal{CN}(\mathbf{0}_N,\underline{\omega}\mathbf{I}_N)$:
        \begin{equation}
            \mathrm{P}_{a\to b}\geq\int_{\mathcal{U}}\frac{\exp\left(-\|\mathbf{t}\|^2/\underline{\omega}\right)}{\pi^{N}|\cov{\bsf{r}|x_a}|}\der\mathbf{t}.
            \label{eq:pep_lower}
        \end{equation}
        
        This lower bound can be expressed in terms of the \acrfull{cdf} of a chi-squared random variable:
        \begin{equation}
            \begin{aligned}
                \mathrm{P}_{a\to b}&\geq|\cov{\bsf{r}|x_a}|^{-1}\underline{\omega}^N\cdot\prob{\|\bsf{t}\|^2\leq|\mathbf{K}|^{-\frac{1}{N}}}{true}\\
                &=|\cov{\bsf{r}|x_a}|^{-1}\underline{\omega}^N\cdot\mathrm{F}_{\chi^2}\left(2\underline{\omega}^{-1}|\mathbf{K}|^{-\frac{1}{N}};2N\right).
            \end{aligned}
        \end{equation}
        If neither $x_a$ nor $x_b$ correspond to the null symbol, the previous bound in the limit of $\alpha\to\infty$ is:
        \begin{equation}
            \lim_{\alpha\to\infty}\mathrm{P}_{a\to b} \geq
            \mathrm{F}_{\chi^2}\left(\frac{2N\log\frac{|x_a|^2}{|x_b|^2}}{\frac{|x_a|^2}{|x_b|^2}-1};2N\right),
        \end{equation}
        which does not vanish for finite $N$ or $\frac{|x_a|^2}{|x_b|^2}$.
        The existence of this lower bound on any \acrshort{pep} is sufficient to prove systems with $M\geq3$ will display a fundamental error floor at high \acrshort{snr}.

        The case for $M=2$ bears some comment.
        In such scenario, the two possible \acrshort{pep}s depend on the null symbol.
        Setting $x_b=0$, we can upper bound $\mathrm{P}_{a\to b}$ in an analogous manner as~\eqref{eq:pep_lower}.
        In this case, we use a wider Gaussian function proportional to the \acrshort{pdf} of $\bsf{t}'\sim\mathcal{CN}(\mathbf{0}_N,\overline{\omega}\id{N})$, with $\overline{\omega}$ being the maximum eigenvalue of $\mathbf{\Omega}$:
        \begin{equation}
            \mathrm{P}_{a\to b}\leq|\cov{\bsf{r}|x_a}|^{-1}\overline{\omega}^N\cdot\mathrm{F}_{\chi^2}\left(2\overline{\omega}^{-1}|\mathbf{K}|^{-\frac{1}{N}};2N\right).
        \end{equation}
        Notice how the upper bound does now vanish as $\alpha\to\infty$:
        \begin{align}
            \lim_{\alpha\to\infty}\mathrm{P}_{a\to b} &\leq \lim_{\alpha\to\infty}\gamma_{\mathrm{MAX}}^N|\mathbf{\Gamma}|^{-1}\cdot\mathrm{F}_{\chi^2}\left(\frac{2\log|\cov{\bsf{r}|x_a}|}{|x_a|^2\gamma_{\mathrm{MAX}}};2N\right)\nonumber\\
            &=\lim_{\alpha\to\infty}\gamma_{\mathrm{MAX}}^N|\mathbf{\Gamma}|^{-1}\cdot\mathrm{F}_{\chi^2}\left(0;2N\right)=0.
        \end{align}

        For the opposite case, in which $x_a=0$, the \acrshort{pep} is obtained as \eqref{eq:pep_int} but now the integration is performed through the region outside of the ellipsoid $\partial\mathcal{D}$.
        With the same change of variable as in \eqref{eq:change}, we map $\mathcal{D}$ to the outside of a ball:
        \begin{equation}
            \mathcal{D}\mapsto\mathcal{V}\triangleq\mathbb{C}^N\setminus\mathcal{U}=\left\{\mathbf{s}\in\mathbb{C}^N:\|\mathbf{s}\|^2>|\mathbf{K}|^{-\frac{1}{N}}\right\}.
        \end{equation}
        The procedure to upper bound this \acrshort{pep} is the same as the one used previously, with the highest eigenvalue of $\mathbf{\Omega}$:
        \begin{align}
            \mathrm{P}_{a\to b}&\leq\int_{\mathcal{V}}\frac{\exp\left(-\|\mathbf{t}\|^2/\overline{\omega}\right)}{\pi^{N}}\der\mathbf{t}=\overline{\omega}^N\cdot\prob{\|\bsf{t}\|^2>|\mathbf{K}|^{-\frac{1}{N}}}{true}\nonumber\\
            &=\overline{\omega}^N\cdot\left(1-\mathrm{F}_{\chi^2}\left(2\overline{\omega}^{-1}|\mathbf{K}|^{-\frac{1}{N}};2N\right)\right).
        \end{align}
        Once again, this upper bound does vanish for increasing \acrshort{snr}:
        \begin{equation}
            \lim_{\alpha\to\infty}\mathrm{P}_{a\to b}\leq1-\lim_{\alpha\to\infty}\mathrm{F}_{\chi^2}\left(2\log\left|\cov{\bsf{r}|x_b}\right|;2N\right)=0.
        \end{equation}
        Therefore, we conclude that systems with $M=2$ will asymptotically be error-free for increasing \acrshort{snr}.
        This completes the proof.\hfill\IEEEQED
        
    \subsection{Mean and Variance of a Quadratic Form in Complex Normal Random Variables} \label{ap:quadratic_form_variance}
        Let $\mathsf{q}=\herm{\bsf{z}}\mathbf{A}\bsf{z} + \mathit{c}$, with $\mathbf{A} = \herm{\mathbf{A}}$ and circularly-symmetric $\bsf{z} \sim \mathcal{CN}\left(\mathbf{0}_N, \cov{\bsf{z}}\right)$.
        Note that by considering an arbitrary $\cov{\bsf{z}}$, $\mathbf{A}$ can be assumed to be diagonal and real without loss of generality.
        The mean of $\mathsf{q}$ is given by:
        \begin{equation}
            \mu_{\mathsf{q}} = \expec{\herm{\bsf{z}}\mathbf{A}\bsf{z} + c}{\bsf{z}}{true} = \trace{\mathbf{A}\cov{\bsf{z}}}{true} + c,
        \end{equation}
        which follows from the cyclic property of the trace.
        
        Using that variance is shift invariant, we obtain:
        \begin{equation} \label{eq:variance_generic_quadratic}
            \var{\mathsf{q}}{\bsf{z}}{true} = \var{\herm{\bsf{z}}\mathbf{A}\bsf{z}}{\bsf{z}}{true}.
        \end{equation}
        Defining $\bsf{z}=\bsf{z}_{\mathrm{R}} + \mathrm{j}\bsf{z}_{\mathrm{I}}$, where $\bsf{z}_{\mathrm{R}} = \real{\bsf{z}}$ and $\bsf{z}_{\mathrm{I}} = \imag{\bsf{z}}$,
        the quadratic form $\herm{\bsf{z}}\mathbf{A}\bsf{z}$ can be expressed as:
        \begin{equation}
            \begin{aligned}
                \herm{\bsf{z}}\mathbf{A}\bsf{z} &= \left(\trans{\bsf{z}_{\mathrm{R}}} - \mathrm{j} \trans{\bsf{z}_{\mathrm{I}}}\right) \mathbf{A} \left(\bsf{z}_{\mathrm{R}} + \mathrm{j} \bsf{z}_{\mathrm{I}}\right)\\
                &= \trans{\bsf{z}_{\mathrm{R}}}\mathbf{A}\bsf{z}_{\mathrm{R}} + \trans{\bsf{z}_{\mathrm{I}}}\mathbf{A}\bsf{z}_{\mathrm{I}} + \mathrm{j}\left(\trans{\bsf{z}_{\mathrm{R}}}\mathbf{A}\bsf{z}_{\mathrm{I}}-\trans{\bsf{z}_{\mathrm{I}}}\mathbf{A}\bsf{z}_{\mathrm{R}}\right)\\
                &= \trans{\bsf{z}_{\mathrm{R}}}\mathbf{A}\bsf{z}_{\mathrm{R}} + \trans{\bsf{z}_{\mathrm{I}}}\mathbf{A}\bsf{z}_{\mathrm{I}}.
            \end{aligned}
        \end{equation}
        Due to $\bsf{z}$ being circularly-symmetric, $\bsf{z}_{\mathrm{R}},\,\bsf{z}_{\mathrm{I}}\sim\mathcal{N}(\mathbf{0}_N, \cov{\bsf{z}}/2)$ are independent, and
        we can compute $\var{\herm{\bsf{z}}\mathbf{A}\bsf{z}}{\bsf{z}}{false}$ as twice the variance of a quadratic form in real random variables~\cite[Ch.~3]{mathai_quadratic_1992}:
        \begin{equation}
            \begin{aligned}
                \var{\herm{\bsf{z}}\mathbf{A}\bsf{z}}{\bsf{z}}{true} &= 2\,\var{\trans{\bsf{z}_{\mathrm{R}}}\mathbf{A}\bsf{z}_{\mathrm{R}}}{\bsf{z}_{\mathrm{R}}}{true} = 4\,\trace{\mathbf{A}\frac{\cov{\bsf{z}}}{2}\mathbf{A}\frac{\cov{\bsf{z}}}{2}}{true}\\ &= \trace{\mathbf{A}\cov{\bsf{z}}\mathbf{A}\cov{\bsf{z}}}{true}.
            \end{aligned}
        \end{equation}
        Substituting in Eq.~\eqref{eq:variance_generic_quadratic}:
        \begin{equation}
            \var{\mathsf{q}}{\bsf{z}}{true} = \trace{\mathbf{A}\cov{\bsf{z}}\mathbf{A}\cov{\bsf{z}}}{true}.
        \end{equation}
        
        Finally, the second-order moment can be easily computed:
        \begin{multline}
            \expec{\mathsf{q}^2}{\bsf{z}}{true} = \var{\mathsf{q}}{\bsf{z}}{true} + \mu_{\mathsf{q}}^2 = \trace{\mathbf{A}\cov{\bsf{z}}\mathbf{A}\cov{\bsf{z}}}{true} + \trace{\mathbf{A}\cov{\bsf{z}}}{true}^2 \\
             + c^2 + 2c\cdot\trace{\mathbf{A}\cov{\bsf{z}}}{true}.
        \end{multline}
    
    \subsection{Energy statistic CRB} \label{ap:crb}
        Given any unbiased estimator $\widehat{\varepsilon}$ of $\varepsilon$,
        the \acrshort{crb} is given by the reciprocal of the Fisher information~\cite[Sec.~3.4]{kay_fundamentals_1993}:
        \begin{equation}
            \var{\widehat{\varepsilon}}{\varepsilon}{true} \geq \frac{1}{\mathrm{J}\left(\varepsilon\right)}.
        \end{equation}
        Under mild conditions, it can be obtained from the log-likelihood function as:
        \begin{equation}
            \mathrm{J}\left(\varepsilon\right)\triangleq-\expec{\frac{\partial^2 l(\varepsilon)}{\partial\varepsilon^2}}{\bsf{r}|\varepsilon}{true}.
        \end{equation}
        In our case, the log-likelihood function is
        \begin{equation} \label{eq:loglikelihood}
            l(\varepsilon) = -\log\left(\pi^{N} \left|\cov{\bsf{r}|\varepsilon}\right|\right) - \herm{\mathbf{r}} \cov{\bsf{r}|\varepsilon}^{-1}\mathbf{r},
        \end{equation}
        with second derivative
        \begin{equation}
            \frac{\partial^2 l(\varepsilon)}{\partial\varepsilon^2} = \sum\limits_{n=1}^{N}\gamma_n^2\frac{\varepsilon\gamma_n+1-2\left|r_n\right|^2}{\left(\varepsilon\gamma_n+1\right)^3}.
        \end{equation}
        The Fisher information is then:
        \begin{equation}
            \begin{aligned}
                \mathrm{J}(\varepsilon) &= -\sum\limits_{n=1}^{N}\gamma_i^2\frac{\varepsilon\gamma_n+1-2\expec{|\mathsf{r}_n|^2}{\bsf{r}|\varepsilon}{true}}{\left(\varepsilon\gamma_n+1\right)^3}\\
                &= \sum\limits_{n=1}^{N}\frac{\gamma_n^2}{\left(\varepsilon\gamma_n+1\right)^2}.
            \end{aligned}
        \end{equation}
        Finally, we can express the \acrshort{crb} in matrix form:
        \begin{equation}
            \var{\widehat{\varepsilon}}{\varepsilon}{true} \geq \frac{1}{\frob{\mathbf{\Gamma}\left(\varepsilon\mathbf{\Gamma}+\id{N}\right)^{-1}}{false}^2} = \frac{1}{\frob{\mathbf{\Gamma}\cov{\bsf{r}|\varepsilon}^{-1}}{false}^2}.
        \end{equation}

    \subsection{Lyapunov CLT} \label{ap:lyapunov}
        Lyapunov \acrshort{clt} is a generalization of Lindeberg--Lévy \acrshort{clt}.
        It states that a sum of a sequence of independent random variables $\{\mathsf{v}_1, \dots, \mathsf{v}_N\}$ with mean $\mu_n$ and variance $\sigma_n^2$ converges in distribution to a normal random variable if the following condition is fulfilled (\ie \textit{Lyapunov's condition}~\cite[Ch.~5]{billingsley_probability_1995}):
        \begin{equation} \label{eq:lyapunov_condition}
            \lim_{N \to \infty} \frac{1}{s_N^{\delta}} \sum\limits_{n=1}^{N} \expec{|\mathsf{v}_n - \mu_n|^{\delta}}{\mathsf{v}_n}{true} = 0,\ \text{for some } \delta > 2,
        \end{equation}
        where $s_N^2 \triangleq \sum_{n=1}^{N} \sigma_n^2$.
        
        In our case, $\mathsf{v}_n \triangleq a_{n} \left|\mathsf{r}_n\right|^2 + c_{n}$ are the summands in~\eqref{eq:estimator_structure} and $N$ is the number of antennas at the receiver.
        Since $\mathsf{r}_n \sim \mathcal{CN}(0, \varepsilon\gamma_n + 1)$, each $\mathsf{v}_n$ follows a shifted exponential distribution with density
        \begin{equation}
            f_{\mathsf{v}_n}(v)=\frac{1}{a_{n}\left(\varepsilon\gamma_n+1\right)} \exp\left(-\frac{v-c_{n}}{a_{n}\left(\varepsilon\gamma_n+1\right)}\right),
        \end{equation}
        defined for $v\in[c_n,\infty)$.
        Its mean is $\mu_n = a_n(\varepsilon\gamma_n+1)+c_n$ and its variance is $\sigma_n^2 = a_n^2(\varepsilon\gamma_n + 1)^2$.
    
        We proceed to verify Lyapunov's condition for $\delta=4$.
        Letting $\mathsf{w}_n = \mathsf{v}_n - \mu_n$, the fourth moment can be easily computed:
        \begin{equation}
            \expec{\mathsf{w}_n^{4}}{\mathsf{w}_n}{true} = \int_{-\sigma_n}^{+\infty} \frac{w^4 \mathrm{e}^{-1}}{\sigma_n} \exp\left(\frac{-w}{\sigma_n}\right) \der w = 9\sigma_n^4.
        \end{equation}
        Substituting in \eqref{eq:lyapunov_condition} and taking into account that $\gamma_n \geq 0$:
        \begin{equation}
            \begin{multlined}
                \lim_{N \to \infty} \frac{1}{s_{N}^{4}} \sum\limits_{n=1}^{N} 9 a_n^4\left(\varepsilon\gamma_n + 1\right)^4\\
                \leq \lim_{N \to \infty} \frac{9}{N} \left(\frac{a_{\mathrm{MAX}} \left(\varepsilon \gamma_{\mathrm{MAX}}+1\right)}{a_{\mathrm{MIN}}\left(\varepsilon \gamma_{\mathrm{MIN}}+1\right)}\right)^4 = 0.
            \end{multlined}
        \end{equation}
        Thus, condition~\eqref{eq:lyapunov_condition} is fulfilled and the \acrshort{clt} can be used.

\bibliographystyle{IEEEtran}
\bibliography{references}

\begin{IEEEbiography}[{\includegraphics[width=1in,height=1.25in,clip,keepaspectratio]{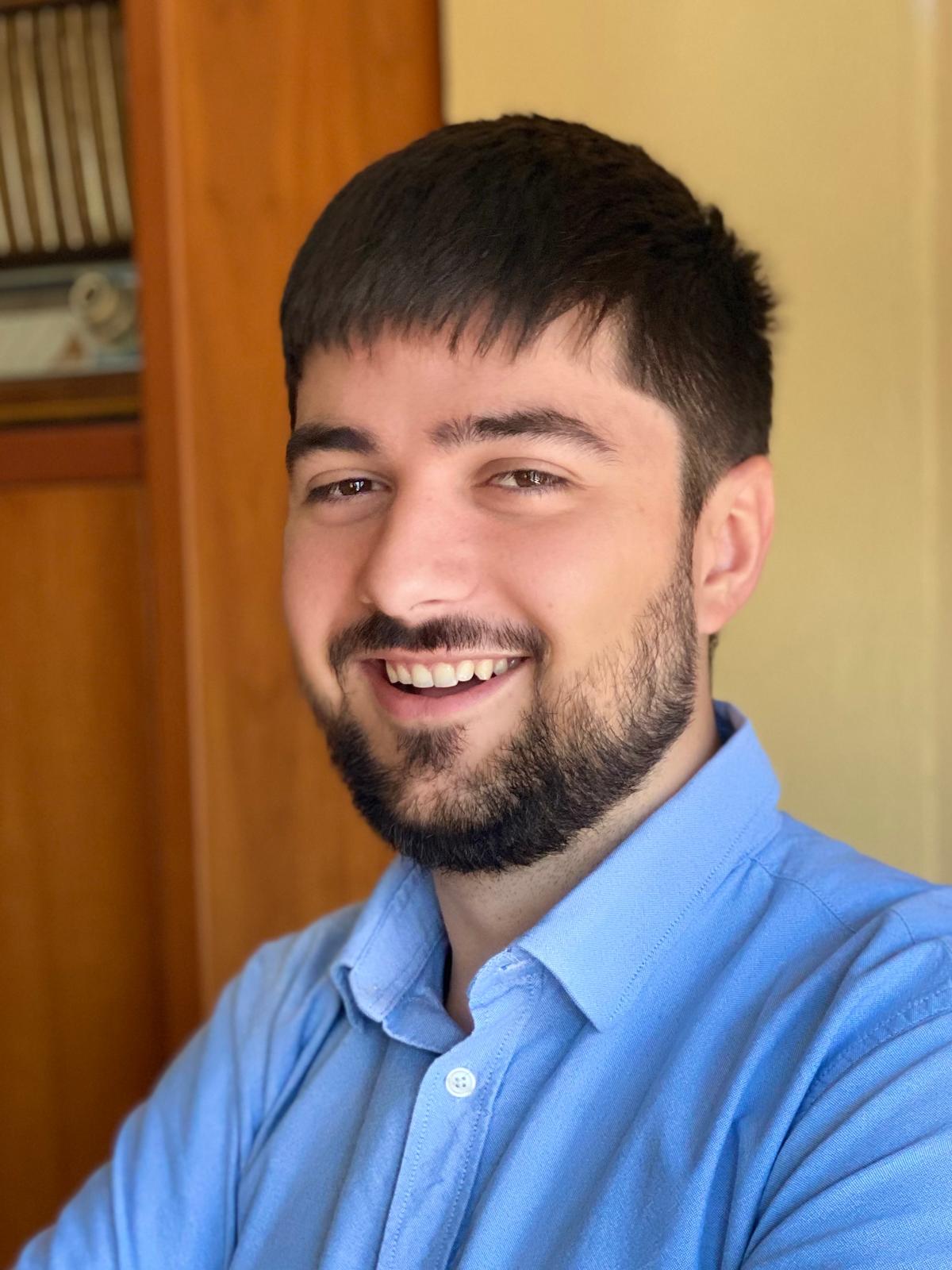}}]{Marc Vilà-Insa}
(Graduate Student Member, IEEE) received the B.Sc. (2019) and M.Sc. (2021) degrees in Telecommunications Engineering from Universitat Politècnica de Catalunya (UPC), Barcelona.
He is currently pursuing a Ph.D. degree in Signal Theory and Communications at the UPC, for which he was awarded with predoctoral grant FI-SDUR in 2022, by the Departament de Recerca i Universitats de la Generalitat de Catalunya.
His areas of expertise are within signal processing for communications.
His research interests encompass topics related to noncoherent communications and massive \acrshort{mimo} systems.
\end{IEEEbiography}

\begin{IEEEbiography}[{\includegraphics[width=1in,height=1.25in,clip,keepaspectratio]{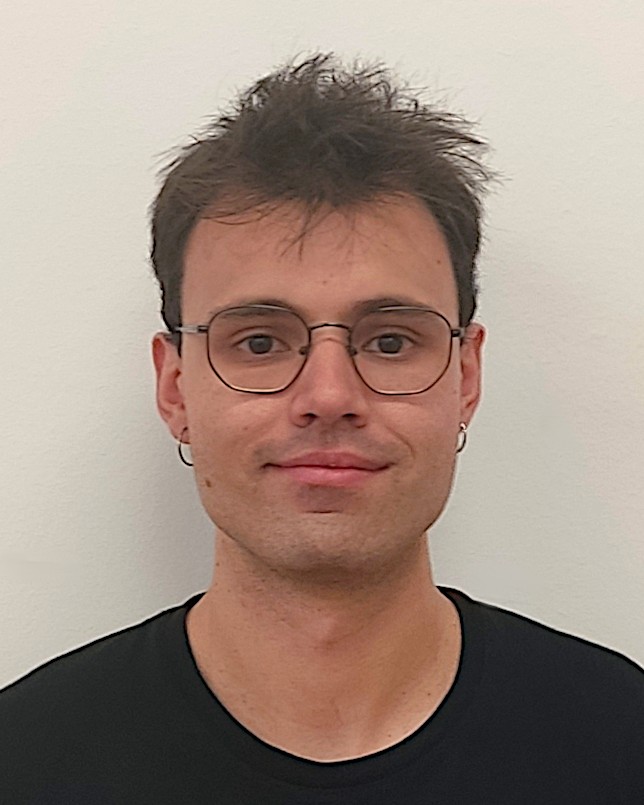}}]{Aniol Martí}
(Graduate Student Member, IEEE) received the B.Sc. and M.Sc. degrees in Telecommunications Engineering from Universitat Politècnica de Catalunya (UPC), Barcelona, in 2021 and 2022, respectively.
From October 2020 to November 2022 he was with DAPCOM Data Services, working on tailored data compression algorithms.
He is currently pursuing a M.Sc. degree in Mathematics at Universidad Nacional de Educación a Distancia (UNED), as well as a Ph.D. degree in Signal Theory and Communications at the UPC.
In 2023, he was awarded with a predoctoral grant (FI-2023 ``Joan Oró'') by the Departament de Recerca i Universitats de la Generalitat de Catalunya.
His research interests are in the areas of wireless communications, statistical signal processing and algebraic statistics.
\end{IEEEbiography}

\begin{IEEEbiography}[{\includegraphics[width=1in,height=1.25in,clip,keepaspectratio]{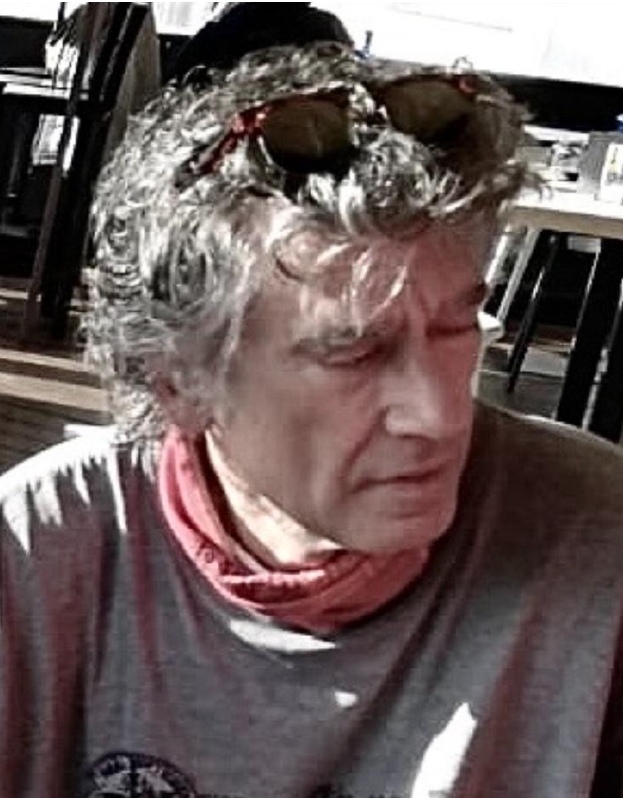}}]{Jaume Riba}
(Senior Member, IEEE) received the M.Sc and Ph.D. degrees in Telecommunications Engineering from the Universitat Politècnica de Catalunya (UPC), Barcelona.
He was then promoted in 1997 to Associate Professor in the same alma mater, teaching subjects related with statistical signal processing and digital communications theories.
He has been regularly involved in research and development programs in the areas of signal processing and satellite communications.
Along with coauthors in the group, Dr. Riba received the 2003 Best Paper Award of the IEEE Signal Processing Society and 2013 Best Paper Award of the IEEE International Conference on Communications.
Having research experience in array processing, synchronization, cyclostationarity, source localization, measures of information, sparsity and noncoherent communications, his current research interest is focused on the interplay between information measures and statistical signal processing principles, looking for interpretability in one field in light of the other.
\end{IEEEbiography}

\begin{IEEEbiography}[{\includegraphics[width=1in,height=1.25in,clip,keepaspectratio]{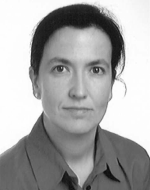}}]{Meritxell Lamarca}
(Member, IEEE) received the M.Sc. and Ph.D. degrees in Telecommunications Engineering from the Universitat Politècnica de Catalunya (UPC), Barcelona, in 1992 and 1997, respectively.
In 1992 she joined the Department of Signal Theory and Communications of the UPC and soon after she became a lecturer.
Since 1997 she has been an Associate Professor at UPC, and from 2009 to 2011 she was on leave at the University of Delaware.
Her general interests are signal processing and digital communications.
She conducts research activities in coding and modulation for wireless communications, with particular emphasis in turbo detection and \acrshort{mimo} systems.
She has been involved in many research projects in wireless and satellite communications under the research programs of the European Union and the European Space Agency. 
\end{IEEEbiography}

\end{document}